\documentclass[12pt,preprint]{aastex}

\def\Bpole{B_{\rm pole}}

\def\fm{\hbox{$.\!\!^{\rm m}$}}

\def\>{$>$}
\def\<{$<$}

\def\simlt{\lower.5ex\hbox{$\; \buildrel < \over \sim \;$}}
\def\simgt{\lower.5ex\hbox{$\; \buildrel > \over \sim \;$}}
\def\ch2{$\chi^{2}$}

\def\ee{\'{e}}

\def\be{\begin{equation}}
\def\ee{\end{equation}}
\def\beq{\begin{eqnarray}}
\def\eeq{\end{eqnarray}}

\def\bB{{\,\mathbf B}}
\def\bE{{\,\mathbf E}}
\def\bj{{\,\mathbf j}}

\def\Rmax{R_{\rm max}}
\def\Rlc{R_{\rm lc}}

\def\rc{r_c}
\def\uc{u_c}
\def\bfe{{\mathbf e}}
\def\fm{f}
\def\fms{f_R}
\def\fmc{f_c}
\def\ta{\psi}
\def\tamax{\psi_{\rm max}}
\def\Etw{E_{\rm tw}}
  \def\Etor{E_{\rm tw}}
\def\Itot{I_{\rm tot}}

\def\tend{t_{\rm end}}
  \def\upc{u_{\rm lc}}
  \def\Iopen{I_{\rm lc}}
  \def\Lopen{L_{\rm lc}}

\def\uesc{u_{\rm esc}}
\def\uf{u_{\star}}

\def\If{I_{\star}}
\def\up{u_m}

\def\BQ{B_Q}
\def\js{j_\star}
\def\V{{\cal V}}
\def\LV{L_{\cal V}}
\def\tV{t_{\cal V}}
\def\Vei{{\cal V}_{ei}}
\def\Vpm{{\cal V}_{\pm}}

\def\tev{t_{\rm ev}}
\def\tg{t_{\rm delay}}

\def\Iref{\hat{I}}

\def\epsradio{\epsilon_{\rm radio}}
\def\tsig{t_\sigma}
\def\XTE{XTE~J1810-197}
\def\AXP{1E~1547.0-5408}

\def\rhoGJ{\rho_{\rm GJ}}

\def\Epar{E_\parallel}

\def\br{{\mathbf r}}

\newbox\grsign \setbox\grsign=\hbox{$>$} \newdimen\grdimen \grdimen=\ht\grsign
\newbox\simlessbox \newbox\simgreatbox \newbox\simpropbox
\setbox\simgreatbox=\hbox{\raise.5ex\hbox{$>$}\llap
     {\lower.5ex\hbox{$\sim$}}}\ht1=\grdimen\dp1=0pt
\setbox\simlessbox=\hbox{\raise.5ex\hbox{$<$}\llap
     {\lower.5ex\hbox{$\sim$}}}\ht2=\grdimen\dp2=0pt
\setbox\simpropbox=\hbox{\raise.5ex\hbox{$\propto$}\llap
     {\lower.5ex\hbox{$\sim$}}}\ht2=\grdimen\dp2=0pt
\def\simgt{\mathrel{\copy\simgreatbox}}
\def\simlt{\mathrel{\copy\simlessbox}}

\begin{document}

\title{Untwisting magnetospheres of neutron stars}

\author{Andrei M. Beloborodov\altaffilmark{1}}
\affil{Physics Department and Columbia Astrophysics Laboratory, 
Columbia University, 538  West 120th Street New York, NY 10027;
amb@phys.columbia.edu}
                                                                                
\altaffiltext{1}{Also at Astro-Space Center of Lebedev Physical
Institute, Profsojuznaja 84/32, Moscow 117810, Russia}

\begin{abstract}
Magnetospheres of neutron stars are anchored in the rigid crust and can 
be twisted by sudden crustal motions (``starquakes''). The twisted 
magnetosphere does not remain static and gradually untwists, dissipating 
magnetic energy and producing radiation. The equation describing this 
evolution is derived, and its solutions are presented. Two distinct regions 
coexist in untwisting magnetospheres: a potential region where
$\nabla\times\bB=0$ (``cavity'') and a current-carrying bundle of field 
lines with $\nabla\times\bB\neq 0$ (``j-bundle''). The cavity has a sharp 
boundary, which expands with time and eventually erases all of the twist. 
In this process, the electric current of the j-bundle is sucked into the star. 
Observational appearance of the untwisting process is discussed.
A hot spot forms at the footprints of the j-bundle. The spot shrinks 
with time toward the magnetic dipole axis, and its luminosity and temperature 
gradually decrease. As the j-bundle shrinks, the amplitude of its twist 
$\ta$ can grow to the maximum possible value $\tamax\sim 1$.
The strong twist near the dipole axis increases the spindown rate of 
the star and can generate a broad beam of radio emission.
The model explains the puzzling behavior of magnetar \XTE~ --- 
a canonical example of magnetospheric evolution following a starquake.
We also discuss implications for other magnetars. The untwisting theory 
suggests that the nonthermal radiation of magnetars is preferentially 
generated on a bundle of extended closed field lines near the dipole axis.
\end{abstract}



\section{Introduction}

Neutron stars are highly conducting and strongly magnetized. 
Their extended magnetospheres are anchored deep in the rigid crust and 
corotate with the star. The magnetosphere is usually assumed to be static 
in the co-rotating frame or evolving very slowly as the star ages.
Electric currents are confined to a narrow bundle of open field lines that 
connect the star to its light cylinder (Goldreich \& Julian 1969). 
The main, closed, part of the magnetosphere is usually assumed to be 
current-free and potential, $\nabla\times\bB=0$.
 
The standard picture of a static potential magnetosphere is reasonable for 
ordinary pulsars, yet apparently it does not describe all neutron stars. 
In particular, neutron stars with ultrastrong fields 
$B\simgt 10^{14}$~G (magnetars) are inferred to have {\it dynamic} 
magnetospheres (see e.g. reviews by Woods \& Thompson 2006; Kaspi 2007;
Mereghetti 2008). Their activity is believed to be caused by crustal
motions relieving internal stresses.\footnote{Well-studied ordinary pulsars 
    (Crab, Vela) also show glitches in their spindown rates, which are 
    associated with sudden crustal deformations.}
In contrast to ordinary pulsars, these objects show huge temporal variations 
in luminosity, spectrum, and spindown rate. Theoretically, stresses are 
expected to build up in the deep crust
as the star ages (e.g. Ruderman 1991; Thompson \& Duncan 1995).
In particular, the large Ampere forces $\bj\times\bB/c$ inside magnetars
can break the crust and shear it in a catastrophic way.\footnote{In some 
    cases, the crust may move plastically.}
Such a starquake twists the magnetic field anchored in the crust, 
creating $\nabla\times\bB\neq 0$ and inducing electric currents in 
the closed magnetosphere (Thompson et al. 2000). The currents are 
approximately force-free and flow along the magnetic field lines, 
$\bj\times\bB=0$. They emerge from the deep crust sheared in the starquake. 

Twisted force-free magnetospheric configurations 
were studied extensively in the context of the solar corona,
and these models can be applied to neutron stars. 
A simple example is the self-similarly twisted dipole.
It was constructed by Wolfson (1995) and applied to neutron stars by 
Thompson, Lyutikov, \& Kulkarni (2002). 
This and similar force-free configurations are {\it magnetostatic} solutions.
A sequence of such configurations may be constructed by changing their 
boundary conditions, i.e. displacing the footpoints of the magnetic field 
lines. If the footpoints freeze, the configuration freezes as well.
At a first glance, this seems to suggest that the implanted twist must 
freeze when the starquake ends, and wait for another starquake.

In fact, the magnetosphere must evolve after the starquake, even though it 
remains anchored in the motionless deep crust. Indeed, energy is continually 
dissipated in the twisted magnetosphere, because  
the twist current $\bj=(c/4\pi)\nabla\times\bB$ is maintained by a 
voltage $\Phi_e\neq 0$ established along the magnetic field lines.
Thompson et al. (2000) estimated voltage $\Phi_e$ assuming that the 
currents are carried by electrons and ions lifted from the 
star's surface against gravity. Beloborodov \& Thompson (2007; hereafter BT07)
found that $\Phi_e$ is regulated by an $e^\pm$ discharge.\footnote{
   The discharge on closed field lines differs from that on open field 
   lines (see Arons 2008 and Beloborodov 2008 for a recent discussion of
   of the polar-cap discharge in ordinary pulsars).
   In both cases, however, the discharge is ultimately driven by the magnetic 
   twist $\nabla\times\bB\neq 0$ that imposes an electric current.
}
The voltage is significant --- comparable to 1~GeV --- and implies a 
modest lifetime of the twist, comparable to one year. 

The untwisting dynamics of the magnetosphere remained, however, unknown.
Usually, resistivity in a plasma leads to diffusion of currents across the 
magnetic field. Voltage $\Phi_e\neq 0$ implies an effective resistivity, 
and one could expect the decaying twist to spread diffusively across 
the magnetosphere. This expectation is incorrect as will be shown below.

The goal of this paper is to develop an electrodynamic theory of twisted 
magnetospheres that describes their evolution. We focus on axially
symmetric configurations. In this case, the twist is created through a 
latitude-dependent azimuthal rotation of the crust. An introductory 
description of twisted magnetic configurations is given in \S~2, 
and their untwisting dynamics is qualitatively discussed in \S~3.
In \S~4 we derive the electrodynamic equation for axisymmetric magnetospheres.

\S~5 presents solutions to the evolution equation and explores the mechanism 
of untwisting. Observational effects of this process are described in \S~6.
\S~7 compares the theory with the recent observations of a starquake in 
the anomalous X-ray pulsar \XTE. \S~8 summarizes the results of the paper 
and discusses implications for magnetars.


\section{Twisted axisymmetric magnetosphere}

In spherical coordinates $r,\theta,\phi$, magnetic field
can be written as the sum of poloidal and toroidal components,
\be
   \bB=\bB_p+\bB_\phi=B_r\hat{\bfe}_r+B_\theta\hat{\bfe}_\theta
                     +B_\phi\hat{\bfe}_\phi,
\ee
where $\hat{\bfe}_r$, $\hat{\bfe}_\theta$, $\hat{\bfe}_\phi$ are unit 
vectors pointing in the $r$, $\theta$, $\phi$ directions.
We assume that the magnetic field is symmetric 
about the polar axis, i.e. $\bB$ does not depend on $\phi$. 
The axisymmetric field can be viewed as a foliation of magnetic
flux surfaces. (Each surface may be obtained by rotating a field line around 
the axis of symmetry.) 

Let $\fm$ be the magnetic flux through a circular
contour of fixed $r=const$ and $\theta=const$. The function $\fm(r,\theta)$ 
is constant on a flux surface. As is usual in plasma physics, we will use 
$\fm$ to label the flux surfaces. Note that flux surfaces extending farther 
from the star have smaller $\fm$, and $\fm=0$ corresponds to the polar axis 
$\theta=0$. We focus on the closed magnetosphere in this paper and neglect
the narrow bundles of open magnetic field lines; effectively, 
rotation of the star is neglected.

Let $R$ be the radius of the star. The magnetic field is force-free outside 
the star:\footnote{Electric fields maintaining the currents in a twisted 
      magnetosphere are relatively weak and do not spoil this approximation.} 
$\bj\times\bB\approx 0$ at $r>R$. If we follow the magnetic field lines 
into the star, significant deviations from the force-free condition appear. 
The $\phi$-component of $\bj\times\bB$ remains, however, small as the crust 
is relatively fragile to axisymmetric azimuthal displacements (which involve 
no compression). This component can be written as $\bj_p\times\bB_p$ where 
$\bj_p$ is the poloidal component of the current density.   
Let $\rc$ be the radius of the lower crust that is strong enough to sustain 
significant azimuthal Ampere forces $\bj_p\times\bB_p/c\neq 0$. 
Outside this radius we assume 
\be
  \bj_p\times\bB_p\approx 0, \qquad   r>\rc.
\ee
A rough estimate $\rc\sim 0.9R$ is sufficient for the purposes of this 
paper (the exact $\rc$ depends on the strength of the magnetic field).
The effective footpoints of the magnetospheric field lines sit at $r=\rc$.
We focus in this paper on the region $r>\rc$ and call it ``force-free''
(in the restricted sense $\bj_p\times\bB_p=0$).

\begin{figure}
\begin{center}
\epsscale{.85}
\plotone{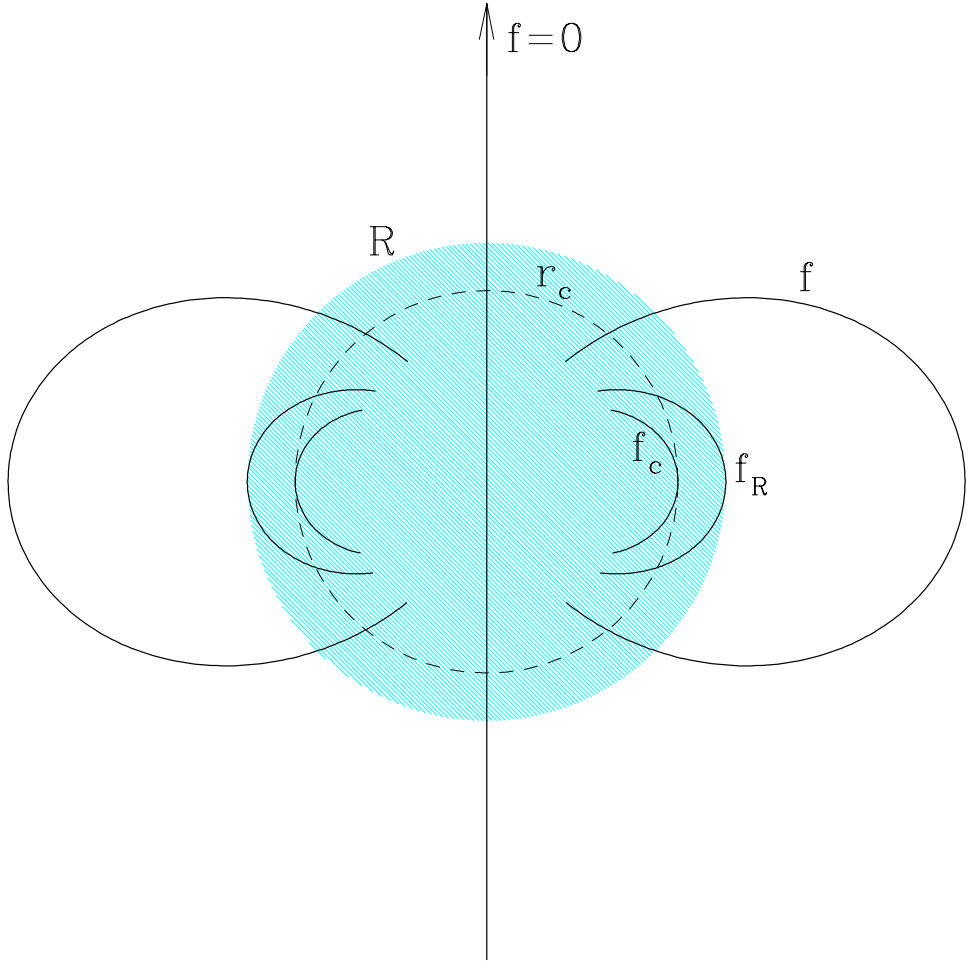}
\caption{
Poloidal cross section of an axisymmetric magnetic configuration. 
The star is the sphere of radius $R$ ({\it shaded}). 
The lower crust is inside radius $\rc$ ({\it dashed circle}).
The figure shows the magnetic axis $\fm=0$, flux surface $\fms$ that 
touches the surface of the star, and flux surface $\fmc$ that touches 
the sphere $\rc$. The magnetosphere is composed of nested closed flux 
surfaces $f<\fms$; one such flux surface is shown in the figure. 
}
\end{center}
\end{figure}

The spheres $r=R$ and $r=\rc$ define two special flux surfaces (Fig.~1):

\noindent
1. Flux surface $\fms$ touches the surface of the star.\footnote{
    Fig.~1 assumes for simplicity that there is only one such flux surface.
    This is the case for a bipolar magnetosphere, with two regions on the 
    star's surface with opposite polarities of the magnetic field (i.e. 
    opposite signs of $B_r$). One can imagine more complicated axisymmetric 
    magnetic configurations with many rings of opposite polarities on the 
    star's surface. The development of electrodynamic theory would be similar 
    in those cases. When calculating the evolution of currents on a given flux 
    surface $\fm$ we would need to know the magnetic field only between the 
    two nearest flux surfaces that touch the surface of the star. 
} 
Flux surfaces $\fm>\fms$ are confined to the star,
and flux surfaces $\fm<\fms$ extend outside the star
and form the magnetosphere. 
$\fms$ represents the total magnetic flux emerging from the star in
the region of positive polarity (where $B_r>0$).

\noindent
2. Flux surface $\fmc$ touches the boundary of the inner crust $r=\rc$.
Flux surfaces $f>\fmc$ are confined to the inner crust and the core of 
the star.

\medskip

A twist is pumped into the force-free region $r>\rc$ when the footpoints 
of field lines at $r=\rc$ are displaced by a starquake.  
The crust is practically incompressible, and hence any axisymmetric 
starquake produces a pure azimuthal displacement (an axisymmetric 
displacement in the $\theta$-direction would imply compression). 

Let us call the footpoints of positive polarity ($B_r>0$) northern.
Consider an initially pure poloidal magnetosphere and suppose a starquake
shifts the northern footpoints of magnetic field lines through angle
$\Delta\phi_n(\fm)$ and southern footpoints through $\Delta\phi_s(\fm)$.
The created twist is described by the relative angular displacement 
$\ta(f)=\Delta\phi_s-\Delta\phi_n=\phi_s-\phi_n$.
This angle can be expressed as an integral along the closed field line:
$\ta=\phi_s-\phi_n$ is accumulated as we move along the field line from 
its northern footpoint at $r=\rc$ to the southern footpoint. 
An infinitesimal displacement $dl$ along the field line corresponds 
to azimuthal displacement $h_\phi\,d\phi=(B_\phi/B)\,dl$
where $h_\phi=r\sin\theta$. Therefore, the twist angle is given by 
\be
\label{eq:ta}
   \ta=\int d\phi=\int_{r>\rc} \frac{B_\phi}{B}\,\frac{dl}{h_\phi},
\ee
where the integral is taken along the field line outside $\rc$.
Creation of $\ta\neq 0$ implies the appearance of toroidal magnetic field
$B_\phi$ in the magnetosphere.

The toroidal field $B_\phi(r,\theta)$ determines the circulation of $\bB$
along the circular contour $r=const$, $\theta=const$.
By Stokes' theorem, it is related to the electric current $I$ flowing through 
the contour,\footnote{The current is maintained through a magnetospheric 
     discharge which fluctuates on a very short (light-crossing) timescale 
     $r/c$ (BT07). We consider the time-average $I$ and treat it as a 
     quasi-steady current. Its evolution caused by resistivity is slow 
     (year timescale) and can be viewed as a slow progression through 
     a sequence of steady states. The displacement current vanishes in a 
     steady state, so $\nabla\times\bB=(4\pi/c)\bj$.} 
\be
\label{eq:Bphi}
   B_\phi=\frac{2I}{cr\sin\theta}.
\ee
$I$ is determined by the poloidal component of the current $\bj_p$, and
$f$ is determined by the poloidal component of the magnetic field $\bB_p$.
At $r>\rc$, the condition $\bj_p\times\bB_p=0$ implies that the poloidal 
currents flow along the poloidal flux surfaces. Therefore, $I$ is a function 
of $\fm$. Note also that the definition of $I$ is similar to that of $\fm$ 
except that $\bB$ is replaced $\bj$, which implies 
\be
\label{eq:j}
  \frac{dI}{df}=\frac{j}{B}.
\ee

Any axisymmetric force-free field outside the star satisfies the 
Grad-Shafranov equation that expresses the condition 
$\bB\times(\nabla\times\bB)=0$ in terms of $I$ and $f$.
Its exact solutions (matching an interior non-force-free solution)
are needed to describe twists with large angles $\ta$. Note that 
configurations with $\ta\gg 1$ are not expected as they are unstable
(Uzdensky 2002 and refs. therein). If the twist grows beyond the instability
threshold $\tamax={\cal O}(1)$, the magnetosphere becomes kink-unstable and 
ejects a closed plasmoid, which prevents the twist growth above $\tamax$.

For moderate 
twists $\ta<1$ one does not have to solve the Grad-Shafranov equation. 
Instead, a simple linear approximation may be sufficient: the configuration 
can be thought of as a linear superposition of an initial non-twisted 
poloidal field $\bB_0$ and a toroidal field $B_\phi<B_0$ that was created 
by the footpoint displacement $\ta$.
The appearance of $B_\phi$ does not affect the poloidal field in the linear 
order --- the poloidal correction to $\bB$ is quadratic in $B_\phi/B$.
The linear approximation may break at large distances from the star 
(Low 1986), however it describes well most of the magnetosphere. 
Wolfson \& Low (1992) found that the relation between $I$ and $f$ obtained
in the linear approximation works well even for twists $\ta\sim 1$.

During the starquake, the twisting motion of the footpoints at $r=\rc$ pumps 
energy into the magnetosphere, which may be released later.
The free energy of twisted force-free configurations was extensively 
studied (e.g. Aly 1984). For linear twists, $B_\phi<B$, the free energy 
simply equals the energy of the toroidal field component $B_\phi$,
\be
    \Etw=\int \frac{B_\phi^2}{8\pi}\,dV,  \qquad \ta<1.
\ee


\section{Fate of the ejected current}

The current $I$ through the magnetosphere is maintained by electric field 
$\Epar\neq 0$ (parallel to $\bB$), which implies Ohmic dissipation of the 
twist energy, $\bE\cdot\bj\neq 0$. Thus, the magnetic field must be 
gradually untwisted (even though it remains anchored in the static 
deep crust),
and eventually the magnetospheric current must vanish. On the other hand, 
the ejected current cannot disappear because it emerges from a static and
almost ideal conductor --- the deep crust $r<\rc$, where the magnetic field 
and electric currents remain unchanged. 
We conclude that Ohmic dissipation must re-direct the ejected poloidal 
current so that it closes below the surface of the star. The current must 
be re-directed {\it across} 
the magnetic flux surfaces, which can happen only in the transition layer 
$r\approx\rc$ between the heavy static conductor and the force-free region. 
Then the current does not penetrate the force-free region on flux surfaces 
$\fm<\fms$ and avoids the magnetospheric dissipation.

It is instructive to consider an idealized problem where the entire star
is a perfect conductor, so that $\Epar\neq 0$ only outside the star.
When the magnetospheric dissipation is completed, $\bj_p=0$
on flux surfaces $f<\fms$ in the region $r>\rc$. The initially ejected
current now flows in a current sheet inside the star (Fig.~2).
The current sheet serves as a screen between the twisted field inside the
star and the untwisted (potential) field in the magnetosphere. 
When the finite resistivity of the crust is taken into account, the state
shown in Figure~2 is not final. The current sheet in the non-ideal conductor 
will acquire a non-zero thickness and gradually spread to the inner flux 
surfaces until the currents reach deeper crust with so high conductivity 
that it can be treated as an ideal conductor on timescales equal to the age 
of the star. The currents that initially emerged during the starquake will 
eventually close deep under the surface of the star.
The tendency of currents to diffuse toward regions of higher conductivity 
was observed in numerical simulations of Ohmic dissipation in neutron star 
crusts (Sang \& Chanmugam 1987).

\begin{figure}
\begin{center}
\epsscale{1.0}
\plottwo{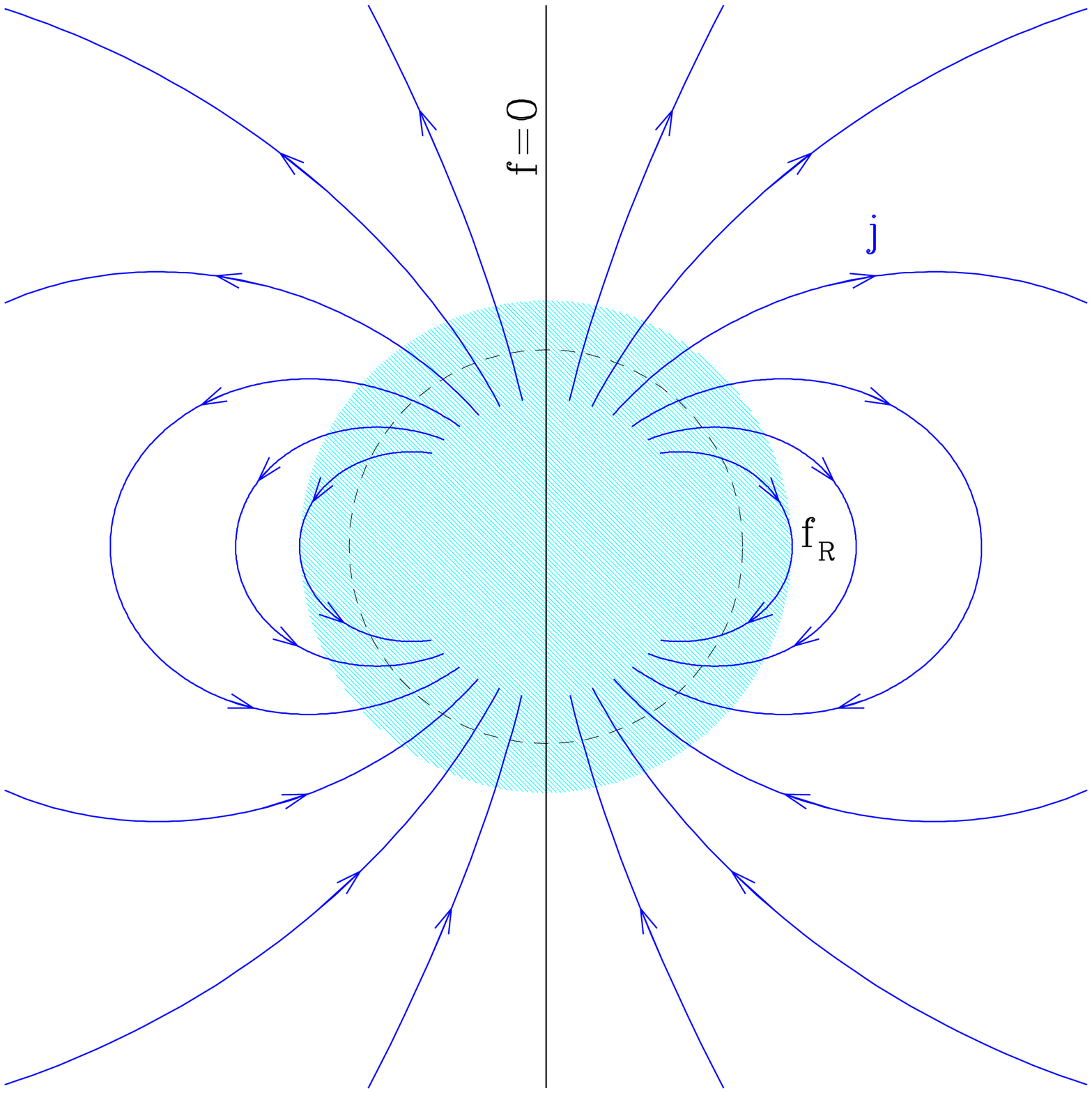}{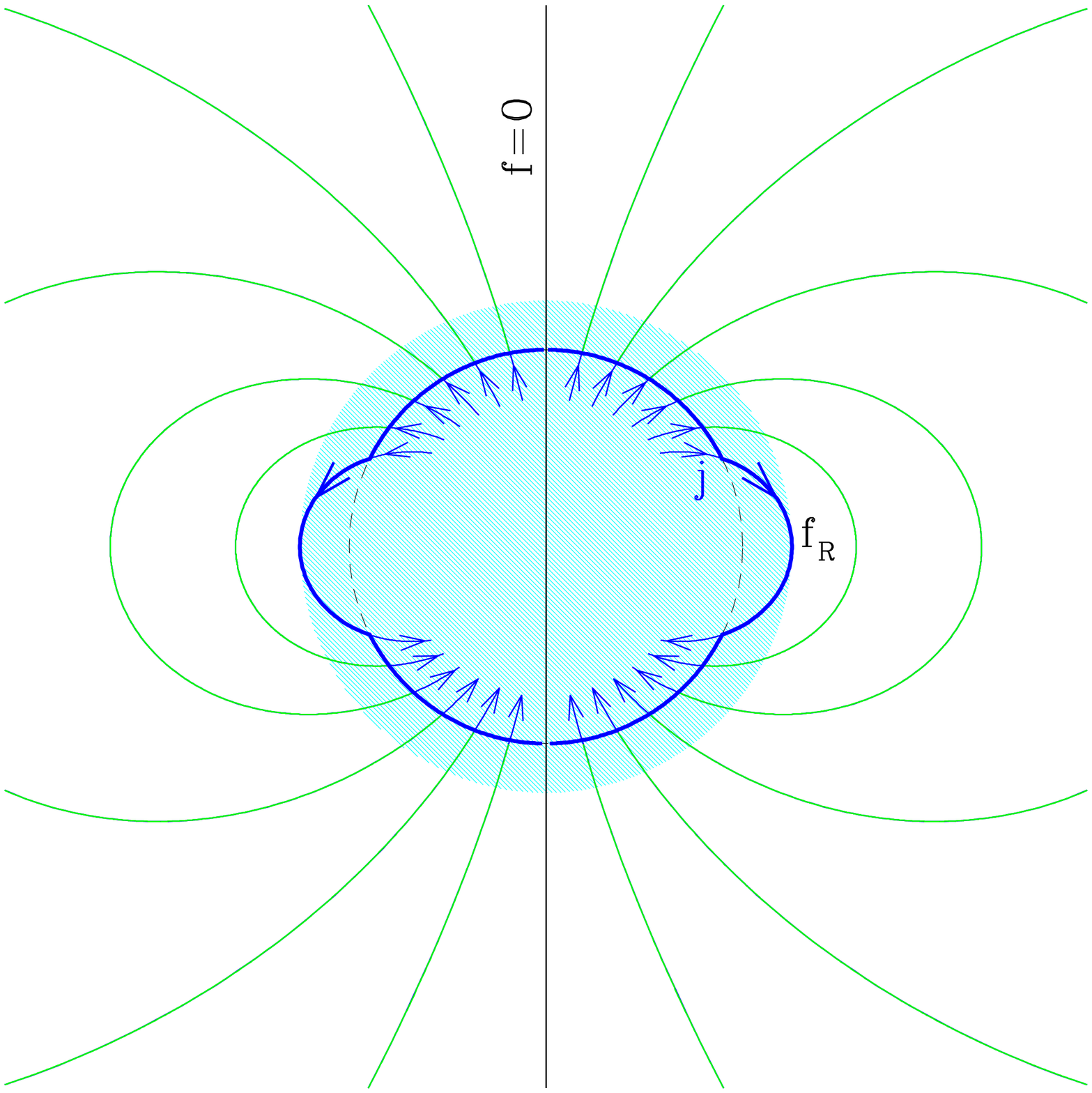}
\caption{
Pattern of electric currents after a global starquake that has twisted the 
entire magnetosphere. Poloidal electric currents are shown by {\it arrows};
they flow along the magnetic flux surfaces outside $r_c$ (dashed circle). 
{\it Left}: initial twisted state. {\it Right}: final untwisted state.
When the magnetospheric dissipation is completed, all currents are
sucked into the star and close below its surface. {\it Thick curve}
shows the poloidal cross section of the current sheet formed inside the
star. It extends from the axis along the sphere $r\approx\rc$, 
across the flux surfaces $\fm<\fms$. The current sheet turns where it 
reaches the flux surface $\fms$, and continues along this flux surface.
}
\end{center}
\end{figure}

In summary, two stages are expected in the evolution of a magnetic twist 
created by a starquake: (1) The current ejected by the starquake
into the magnetosphere is gradually drawn into the star. Most of the twist 
energy is released at this stage. (2) The current spreads into deeper 
layers of the star and eventually collects near the highly conducting inner 
crust. The second, subsurface untwisting stage is much slower because the 
resistivity inside the star is smaller than the effective resistivity of 
the magnetosphere. We shall focus below on the faster first stage and treat 
the star as an ideal conductor. 


\section{Twist evolution equation}
\label{sc:ev_eq}

\subsection{Resistive evolution}
\label{sc:evol}

The evolution of magnetic field is related to 
electric field $\bE$ according to the induction equation,
\be
   \frac{1}{c}\,\frac{\partial\bB}{\partial t}=-\nabla\times\bE.
\ee
We will express $\nabla\times\bE$ in curvilinear coordinates $q^i$
defined as follows. Let us label magnetic field lines on a flux 
surface $\fm$ by the azimuthal angle $\phi_0$ of their northern footpoints
at $r=\rc$ (northern footpoints have $B_r>0$).
Thus, the set of all field lines is parameterized by two coordinates
$\fm$ and $\phi_0$. Let $s$ be a parameter running along the field line
(increasing in the direction of $\bB$). 
The parameter $s$ may be chosen, e.g. equal to length $l$ measured along 
the field line from its northern footpoint (then $|\bfe_s|=1$).
The coordinate system $q^i=(s,\fm,\phi_0)$ covers the entire magnetic field 
that passes through the sphere $r=r_c$ and emerges in the force-free region. 

The electric field can be written in components in the new coordinate basis,
\be
   \bE=\sum_{i=1}^3 E^i\bfe_i, \qquad 
   \bfe_i=\frac{\partial{\mathbf r}}{\partial q^i}.
\ee
The length of the basis vectors will be denoted by $h_i=|\bfe_i|$. Note that
\be
   \bfe_3\equiv\left.\frac{\partial{\mathbf r}}{\partial \phi_0}\right|_{f,s}
 =\left.\frac{\partial{\mathbf r}}{\partial \phi}\right|_{r,\theta}=\bfe_\phi
  =h_\phi\hat{\bfe}_\phi.
\ee
We will need the determinant $g$ of metric $g_{ik}=\bfe_i\cdot\bfe_k$, 
which is given by $\sqrt{g}=\bfe_s\cdot(\bfe_{\fm}\times\bfe_\phi)$. 
Consider an infinitesimal axisymmetric ring perpendicular to the poloidal
component of $\bB$.
Its surface element is $\bfe_{\fm}\times\bfe_\phi\,d\fm\,d\phi$, and
the area of the ring is $2\pi\,|\bfe_{\fm}\times\bfe_\phi|d\fm$. 
The magnetic flux through the ring, $d\fm$, is
\be
 d\fm=2\pi\bB\cdot\left(\bfe_{\fm}\times\bfe_\phi\right)\;d\fm,
\ee
which implies the identity 
$2\pi\bB\cdot\left(\bfe_{\fm}\times\bfe_\phi\right)=1$.
We substitute $\bB=B\bfe_s/h_s$ and find 
\be
\label{eq:g}
   \sqrt{g}=\frac{h_s}{2\pi B}.
\ee

The general expression for $\nabla\times\bE$ in curvilinear coordinates is
\be
   \left(\nabla\times\bE\right)^i=\sum_{j,k=1}^{3}
    \frac{\epsilon^{ijk}}{\sqrt{g}}\,\frac{\partial{E_k}}{\partial q^j},
\ee
where $\epsilon^{ijk}$ is Levi-Civita symbol and $E_i=\bfe_i\cdot\bE$ are 
the covariant components of the electric field in coordinate system $q^i$.
Using $\bfe_i=h_i\hat{\bfe}_i$, we obtain the azimuthal component of 
$\nabla\times\bE$ in the normalized basis $\hat{\bfe}_i$ and find
\be
   \frac{1}{c}\,\frac{\partial B_\phi}{\partial t}=
  h_\phi\,\frac{2\pi B}{h_s}\,
 \left(\frac{\partial E_s}{\partial f}-\frac{\partial E_f}{\partial s}\right).
\ee
Let us divide both sides of this equation by $2\pi B h_\phi/h_s$ and 
integrate it over $s$ along an entire closed field line (including its 
part at $r<\rc$). Then the second term on the right-hand side disappears, 
and we get
\be
\label{eq:bbb}
  \frac{1}{2\pi c}\,\oint\frac{\partial B_\phi}{\partial t}\,
  \frac{h_s\,ds}{h_\phi B}=\frac{\partial}{\partial f}\oint E_s\,ds.
\ee 
The integral on the right-hand side is the net voltage 
induced along the magnetic field line,\footnote{This voltage is not 
electrostatic and does not vanish for a closed contour. It is the 
self-induction voltage of the gradually decaying twist, see \S~2 in BT07.}
\be
\label{eq:Phi_e}
   \Phi_e=\oint E_s\,ds=\oint \Epar\,dl,
\ee
where $\Epar=\bE\cdot\hat{\bfe}_s$ and $dl=h_sds$ is the length element 
along the field line. Equation~(\ref{eq:bbb}) shows that the twist evolution 
is controlled by the longitudinal voltage, as expected. 
Using $B_\phi=2I(f,t)/ch_\phi$ (eq.~\ref{eq:Bphi}), we rewrite this equation 
as
\be
\label{eq:bb}
  \frac{1}{2\pi c}\,\oint \frac{1}{I}\,\frac{\partial I}{\partial t}\,
  \frac{B_\phi \,dl}{B\,h_\phi}=\frac{\partial\Phi_e}{\partial f}.
\ee 
Note that $\partial B_\phi/\partial t=0$ and $\partial I/\partial t=0$
at $r<r_c$ since the magnetic 
field remains static inside the static ideally conducting inner region. 
$\partial I/\partial t$ jumps from 0 to its value in the force-free region
at $r\approx \rc$, in a transition layer of thickness $\Delta\ll\rc$.
The contribution from this layer to the integral on the left-hand side 
of equation~(\ref{eq:bb}) is small and we neglect it. Then, effectively, 
the integral is taken only along the force-free part of the field line 
at $r>r_c$. The current $I$ and its time derivative 
$\partial I/\partial t$ are constant along this part of the field line. 
Then, using the expression for the twist angle (eq.~\ref{eq:ta}), we find
\be
\label{eq:ev2}
  \frac{\ta}{2\pi cI}\,\frac{\partial I}{\partial t}\, 
  =\frac{\partial\Phi_e}{\partial f}.
\ee
Equation~(\ref{eq:ev2}) describes the evolution of $I(f,t)$ and the 
corresponding $\ta(f,t)$. The twist is changing with time as the 
magnetic field lines gradually slip in the resistive magnetosphere and 
connect new footpoints with different $\phi_s-\phi_n$. Thus, $\ta$
is changing despite the fact that the magnetosphere remains anchored in 
the static deep crust.

\subsection{Linear twists}
\label{sc:linear}

For small twists $\ta<1$, the magnetic field can be written as
$\bB\approx \bB_0+\bB_\phi(t)$ (\S~2), where poloidal field $\bB_p=\bB_0$ 
remains static and $B_\phi<B_0$. Then 
$B=B_0[1+{\cal O}(B_\phi^2/B^2)]\approx const$ 
and the twist evolution equation~(\ref{eq:bb}) simplifies to
\be
\label{eq:ev3}
  \frac{1}{2\pi c}\,\frac{\partial \ta}{\partial t}
  =\frac{\partial\Phi_e}{\partial f}.
\ee
This approximate equation
quickly becomes accurate for $B_\phi\ll B$: its error is decreasing 
as $(B_\phi/B)^2$. In the following sections we will use 
equation~(\ref{eq:ev3}) to study the evolution of the twist amplitude $\ta$.
The exact evolution equation~(\ref{eq:ev2}) would have to be used
when the effect of the twist on $\bB_p$ is of interest (e.g. for 
calculations of the spindown rate of the star), 

The linearized description of twisted configurations is useful for the 
first calculations of the resistive untwisting. However, the limitations of 
this approximation should be kept in mind. The linearized description may 
fail at large distances from the star (Low 1986). Besides, the force-free 
configuration may be unable to smoothly adjust to the growing twist. 
Sudden relaxation to a new topological configuration is possible,
with partial opening of the field lines and the loss of connectivity 
between the footpoints. The twist evolution equation derived above 
(linear or nonlinear) does not describe such transitions.

\subsection{Energy conservation law}
\label{sc:energy}

The energy of a linear twist $B_\phi<B_0$ is given by
\be
   \Etor\equiv\int_{r>\rc} \frac{B_\phi^2}{8\pi}\, dV =\frac{1}{8\pi}
       \int\int\int\frac{2I}{ch_\phi}\,B_\phi\,\sqrt{g}\,ds\,df d\phi
   =\frac{1}{4\pi c}\int_0^{\fmc} I(f)\,\ta(f)\,df.
\label{eq:en0}
\ee
Differentiating with respect to time and using equations~(\ref{eq:ev2})
and (\ref{eq:ev3}), we get
\be
  \frac{d\Etor}{dt}=\int_0^{\fmc} I\,\frac{\partial\Phi_e}{\partial f}\,df.
\ee
Integrating by parts and taking into account that $I(0)=0$ and 
$\Phi_e(\fmc)=0$ ($\Epar=0$ on flux surfaces confined to the perfect 
conductor), we find
\be
\label{eq:en}
  \frac{d\Etor}{dt} =-\int \Phi_e \,dI.
\ee
The right-hand side represents the net Ohmic losses. 
$\Etor$ is the free energy of the twist that is gradually dissipated 
as the magnetic field evolves according to equation~(\ref{eq:ev3}).


\subsection{Magnetosphere with moving footpoints}

The evolution equation derived above assumes that the magnetosphere is
anchored in the deep {\it static} crust following a starquake.
It is straightforward to generalize equation~(\ref{eq:ev3}) for
magnetospheres with moving footpoints,
\be
\label{eq:ev4}
  \frac{\partial \ta}{\partial t}
  =2\pi c\,\frac{\partial\Phi_e}{\partial f}+\omega(f,t),
\ee
where $\omega=d\phi_s/dt-d\phi_n/dt$ is the differential angular velocity 
of the northern and southern footpoints of the magnetic field lines.
A crust that remains motionless at all times except a sudden starquake
at $t=0$ is described by
\be
\label{eq:quake}
  \omega(f,t)=\ta_0(f)\,\delta(t),
\ee
where $\delta(t)$ is the Dirac function and $\ta_0$ is the amplitude of 
the twist imparted by the starquake.
The impulsive twisting is a good approximation for recurring
starquakes if the time between subsequent starquakes is longer than the 
timescale of untwisting. In the opposite limit, the crust is frequently 
deformed my mini-starquakes or moves plastically.
Then the footpoint motion can be described by a continuous function 
$\omega(f,t)$. In this paper, we focus on the case of impulsive twisting 
described by equation~(\ref{eq:quake}).

\subsection{Twisted dipole}
\label{sc:dipole}

For the study of the untwisting mechanism in the next section, 
it is useful to consider a concrete simple magnetic configuration.
We will consider a dipole field with the symmetry axis passing through 
the center of the star. Let $\vec\mu$ be the dipole moment.
The poloidal flux function for the dipole is given by (Appendix~A),
\be
\label{eq:fm}
   \fm(r,\theta)=2\pi\mu\,\frac{\sin^2\theta}{r}
                =\frac{2\pi \mu}{\Rmax},
\ee
where $\Rmax=r/\sin^2\theta$ is the maximum radius reached by the flux
surface passing through given $r,\theta$.
Hereafter, instead of $\fm$ or $\Rmax$ we will label flux surfaces by the
dimensionless coordinate
\be
\label{eq:u}
    u\equiv\frac{\fm}{\fms}=\frac{R}{\Rmax}.
\ee
The last flux surface in the force-free region (marginally emerging from
the inner crust: $\Rmax=\rc$) has $u=\uc=R/\rc\approx 1.1$.
The region $0<u<1$ corresponds to the magnetospheric flux surfaces.
In this region, $u=\sin^2\theta_1$, where $\theta_1$ is the polar angle of 
the northern footprint of the field line on the star surface $r=R$.
                                                                                
Suppose now that the field has been twisted by a starquake that was 
symmetric about the dipole axis and resulted in a differential rotation 
of the crust through angle $\ta(u)$ (\S~2).  The poloidal current $I$ and 
twist angle $\ta$ will be viewed below as functions of $u$ and time $t$.
The following relation holds between $\ta$ and $I$
\be
\label{eq:ta2}
  \ta=\frac{4R^2I}{c\mu u^2}\sqrt{1-\frac{u}{\uc}},
\ee
(see Appendix~A). The twist evolution equation~(\ref{eq:ev3}) becomes,
\be
\label{eq:evd1}
   \frac{\partial\ta}{\partial t}
  =\frac{cR}{\mu}\,\frac{\partial{\Phi_e}}{{\partial u}},
\ee
or
\be
\label{eq:evd2}
  \frac{\partial I}{\partial t}
  =\frac{c^2u^2}{4R\sqrt{1-u/\uc}}\,\frac{\partial{\Phi_e}}{{\partial u}}
\ee

The current density in the twisted dipole magnetosphere
is given by (cf. eq.~\ref{eq:j}),
\be
\label{eq:jd}
   j=\frac{RB}{2\pi\mu}\frac{\partial I}{\partial u}.
\ee


\section{Mechanism of untwisting}

This section will explore the mechanism of untwisting for twists created 
by axisymmetric starquakes in the dipole magnetosphere (\S~\ref{sc:dipole}). 
Their evolution is described by equation~(\ref{eq:evd1}). Before this 
equation can be solved, the voltage $\Phi_e$ must be specified.

\subsection{Twist evolution in a medium with fixed conductivity}
\label{sc:sigma}

Consider first what would happen if the magnetosphere was filled 
by a medium with a fixed conductivity $\sigma$; we will assume in this 
toy model $\sigma(\br)=const$. Then the current density is related to 
$\Epar$ by Ohm's law $j=\sigma\Epar$. The star will be modeled as a perfect 
conductor, so $\Epar=0$ is assumed inside the star. Then the voltage 
$\Phi_e$ along a magnetospheric field line is given by 
\be
   \Phi_e=\int_{r>R} \frac{j}{\sigma}\,dl
         =\frac{\sqrt{1-u}}{\pi R\sigma}\frac{\partial I}{\partial u}.
\ee
(The integral has been calculated using eqs.~\ref{eq:jd} and \ref{eq:int}.)
Substitution of this result to the twist evolution equation~(\ref{eq:evd2}) 
yields the equation for $I(u,t)$,
\be
\label{eq:ev_sig}
  \frac{\partial I}{\partial t}
 =\frac{c^2u^2}{4\pi\sigma R^2}\left(1-\frac{u}{\uc}\right)^{-1/2}
   \frac{\partial}{\partial u}
   \left(\sqrt{1-u}\,\frac{\partial I}{\partial u}\right).
\label{eq:diff}
\ee
Given an initial twist with current function $I(u,0)$, one can calculate 
the evolution of $I(u,t)$ by solving this differential equation with
two boundary conditions: $I(0)=0$ and $I(1)=const$. The latter condition 
is valid as long as the perfect-conductor approximation is used for the star.

Equation~(\ref{eq:diff}) is of diffusion type. It has a special feature: 
the effective diffusion coefficient is proportional to $\sqrt{1-u}$ and 
vanishes at $u=1$. This fact allows one to qualitatively understand the 
evolution of $I(u,t)$ before solving the equation numerically.
It is instructive to consider the analogous problem of particle diffusion
with a position-dependent diffusion coefficient $D(u)$ that vanishes at 
the boundary $u=1$. The vanishing of $D$ means that particles ``stick'' to 
the boundary. With time, more and more particles get stuck, and eventually 
the particle density vanishes everywhere except at the boundary.
The magnetospheric current behaves in a similar way. Far from the boundary,
it simply spreads diffusively: 
$\partial I/\partial t=const\,u^2\,\partial^2I/\partial u^2$ at $u\ll 1$.
At the same time, near the boundary $u=1$, the current is sucked toward 
$u=1$. The current keeps accumulating at $u=1$ until $I=0$ at all $u<1$.

For illustration, we solved numerically equation~(\ref{eq:diff}) for 
a twist that initially has a uniform amplitude $\ta_0(u)=0.2$. 
The corresponding initial current function $I(u,0)$ and 
$\partial I/\partial u\propto j/B$ are shown in Figure~3. 
The evolution of this twist is shown in Figure~4.
The characteristic diffusion timescale is $\tsig=R^2c^2/4\pi\sigma$, 
and we express time in units of $\tsig$.
As expected, the current tends to spread to smaller $u$ and, at 
the same time, it is quickly drawn into the current sheet at $u=1$.
The sum of the currents flowing in the magnetopshere and in the current 
sheet, $\Itot$, remains constant.
Our numerical model assumes $V=0$ at $u>1$ (inside the star),
which allows the current sheet to persist at $u=1$. A real star has
a finite conductivity and the current sheet will slowly spread into
the star (toward larger $u>1$). 

Note that the twist amplitude $\ta$ grows near $u=0$, because 
$\partial\Phi_e/\partial u>0$ in this region. 
A strongly twisted bundle develops near the magnetic axis, however, its 
thickness shrinks with time, so its net current decreases. At late times, 
the twist growth near the axis enters a self-similar regime. 
The amplitude peak $\ta_{\rm peak}$ would grow indefinitely in the limit 
$t\rightarrow\infty$, if the twist remained stable. In fact, the growth 
must be stopped by the MHD instability expected at $\ta\simgt 1$. 
An upper limit on $\ta$ is set by the field-line opening,
which may give a partially opened configuration with a lower energy
(Wolfson \& Low 1992). Besides, the strongly twisted configuration is
prone to kink instability, which is likely to reconnect part of the twisted
region away from the star. These dynamic processes are not described by our 
model. We shall assume that they keep $\ta$ below $\tamax={\cal O}(1)$.

\begin{figure}
\begin{center}
\epsscale{.80}
\plotone{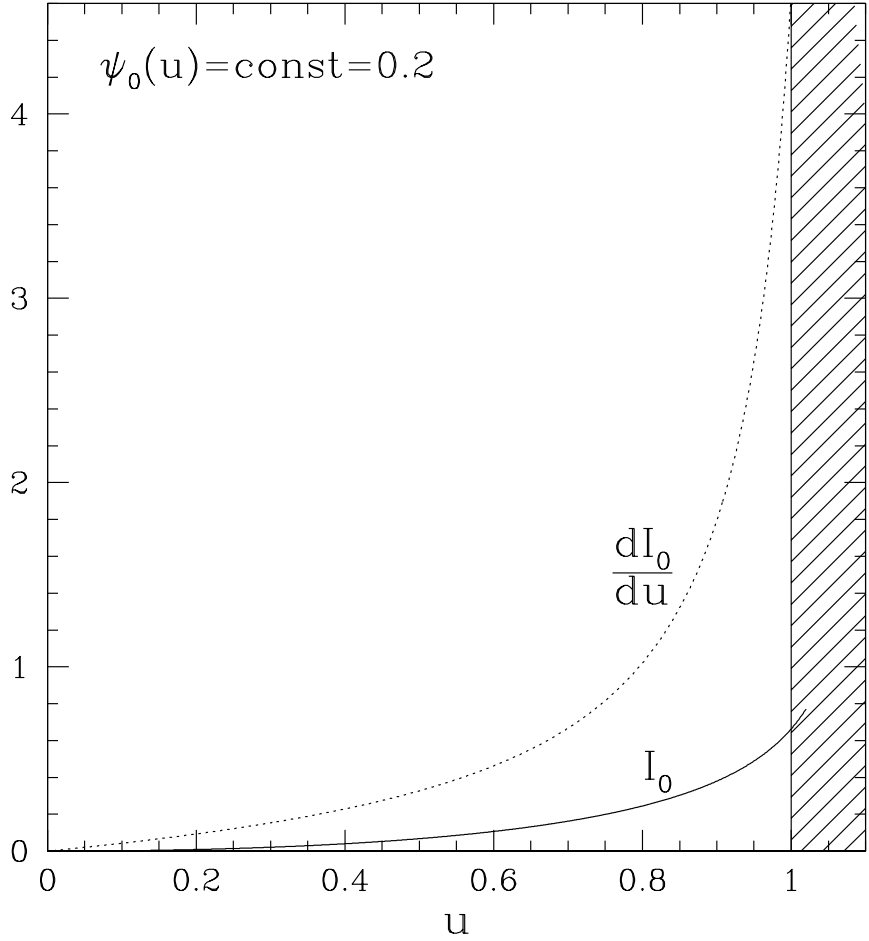}
\caption{
A uniform twist created by a global starquake involving the entire 
magnetosphere $0<u<1$. The twist amplitude in this example is 
$\ta_0(u)=const=0.2$.
The figure shows the corresponding current function $I_0(u)=I(u,0)$ and 
its derivative $dI_0/du$, which is proportional to current density $j$ 
(see eq.~\ref{eq:jd}). Current $I$ is measured in units of $\Iref=c\mu/4R^2$,
where $\mu$ and $R$ are the magnetic dipole moment and the radius of the star.
The coordinate $u=\fm/\fms=R/\Rmax$ labels flux surfaces (see eq.~\ref{eq:u} 
and Fig.~1); $\Rmax$ is the maximum radius reached by the flux surface. 
$u=0$ on the magnetic axis ($\Rmax\rightarrow\infty$ on the axis). 
For magnetospheric flux surfaces $u<1$ and $u=\sin^2\theta_1$ where 
$\theta_1$ is the polar angle of the northern footprints of the flux surface 
on the star. Flux surfaces with $u>1$ close inside the star; this region is 
shaded in the figure.
}
\end{center}
\end{figure}
\begin{figure}
\begin{center}
\epsscale{.9}
\plotone{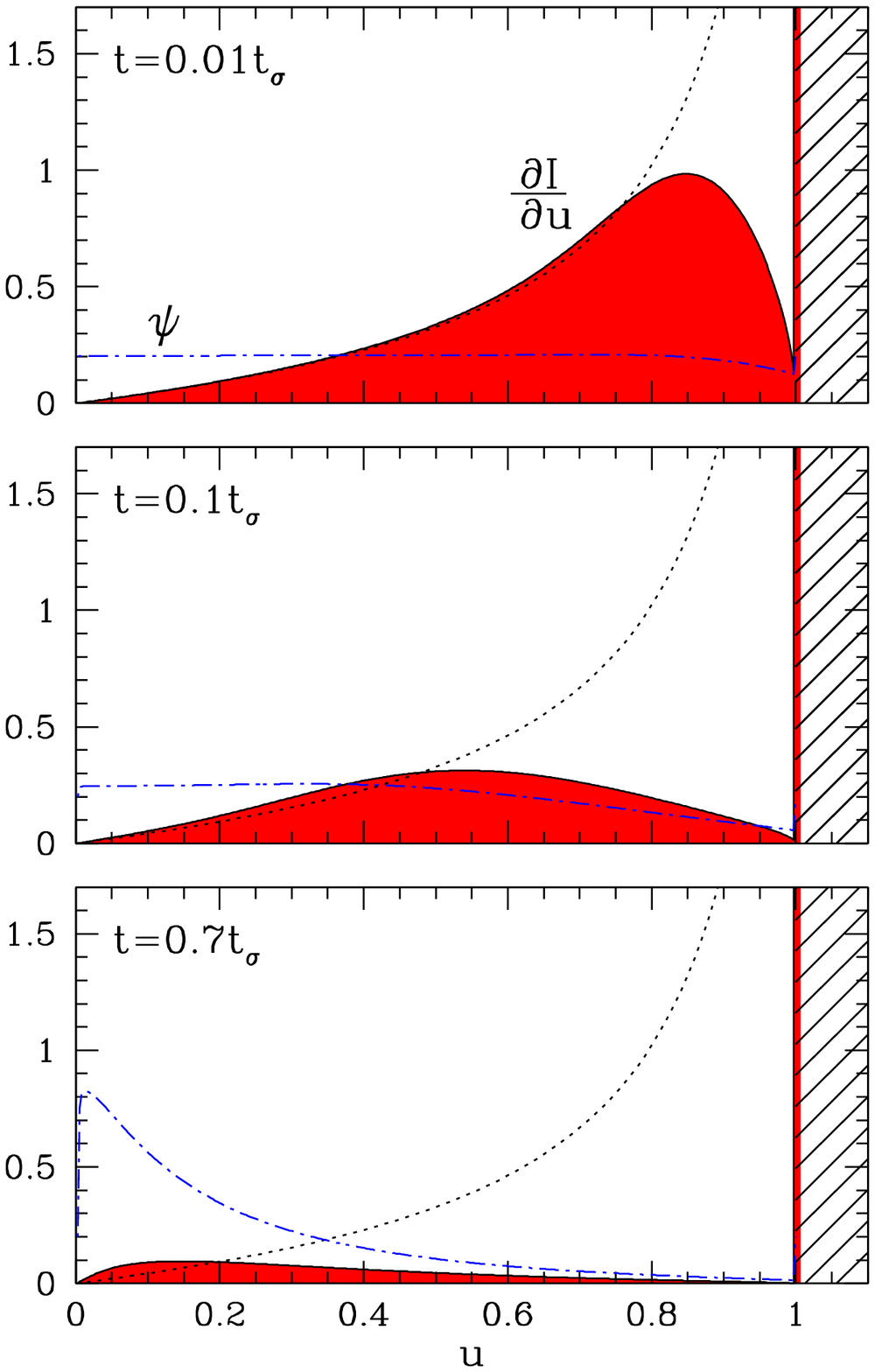}
\caption{
Evolution of the twist (with the initial configuration shown in Fig.~3),
calculated under the assumption that the magnetosphere is filled with a 
medium of a fixed conductivity $\sigma$, and the star is an ideal conductor.
Three panels show the twist at times $t/\tsig=0.01$, 0.1, and 0.7, where 
$\tsig=R^2c^2/4\pi\sigma$. 
{\it Solid curve} shows the current distribution $\partial I(u,t)/\partial u$;
$I$ in the figure is measured in units of $\Iref=c\mu/4R^2$.
The initial $dI_0(u)/du$ is shown by {\it dotted curve} (from Fig.~3). 
{\it Shaded area} represents the magnetospheric current $\If$ and the current 
sheet $I_1$ at $u=1$. The total current $\Itot=\If+I_1$ is conserved.
$I_1$ grows with time, and the magnetospheric current $\If$ decreases and 
spreads toward the magnetic axis $u=0$. {\it Dashed-dotted curve} shows 
the twist amplitude $\ta$. It grows near the axis. 
}
\end{center}
\end{figure}


\subsection{Threshold voltage}
\label{sc:voltage}

The toy model discussed in \S~5.1 is deficient: the real magnetospheres 
of neutron stars are not filled by a medium of a fixed conductivity. 
Instead, the magnetospheric currents are maintained through a discharge
with a \emph{threshold voltage} that is approximately the same for any 
current density $j\neq 0$ (or, more precisely, for any $j$ exceeding a 
small value $\js$). The threshold nature of the magnetospheric voltage 
changes the evolution of the twist. Remarkably, it simplifies the solution 
of the twist evolution equation: the partial differential equation will 
be reduced to an ordinary differential equation.

We denote the threshold voltage by $\V$. Once $\Phi_e$ reaches $\V$, 
copious particle supply is available to carry any large current. 
The particle supply and voltage regulation in magnetars was studied in BT07. 
In principle, there are two sources of particles: 
(1) the surface of the star and (2) $e^\pm$ creation in the magnetosphere.
The current is carried by charges of both signs, since the net charge
density must be nearly zero to avoid huge voltages. If no $e^\pm$ are 
created, electrons and ions must be lifted from the star's surface 
(if ions are available in an atmospheric layer atop the solid crust). 
Maintaining a flow of ions along a magnetic loop requires a minimum voltage,
\be
\label{eq:Vei}
  e\Vei=\frac{GMm_i}{R}-\frac{GMm_i}{\Rmax},
\ee
where $\Rmax$ is the maximum radius reached by the loop and $m_i$ is the 
ion mass. This voltage is $\sim 0.2m_ic^2$ for loops with $\Rmax\gg R$. 
The plasma is lifted into the loop by the self-induction electric field 
(BT07). The exact solution was obtained for this process in the 
one-dimensional circuit model. It shows that voltage keeps growing even 
after reaching $\Vei$, and the circuit evolves toward a global double-layer 
configuration. This result may not hold in the complete 3D model that 
includes the excitation of transverse waves in the magnetosphere. 
Nevertheless, it suggests that lifting plasma from the surface is not 
the ultimate regulator of the voltage.

A robust mechanism for limiting the voltage is the $e^\pm$ discharge. 
If $\Phi_e$ exceeds a certain threshold $\Vpm$, the exponential runaway 
of pair creation occurs (cf. Fig.~5 in BT07), and the $e^\pm$ pairs screen 
$\Epar$.  In a magnetar magnetosphere, the $e^\pm$ avalanche is triggered 
when the accelerated electrons resonantly scatter X-rays streaming from the 
star, and the scattered photons convert to $e^\pm$ off the magnetic field.
The threshold for the discharge is given by 
\be
\label{eq:Vepm}
  e\Vpm\sim \gamma_{\rm res} m_ec^2\sim \frac{ceB}{\omega_X}
  \approx 1\left(\frac{B}{10^{14}~{\rm G}}\right)
           \left(\frac{\omega_X}{10^{18}{\rm~Hz}}\right) {\rm~GeV},
\ee
where $\omega_X$ is the typical frequency of target X-rays 
and $\gamma_{\rm res}\sim (B/\BQ)(m_ec^2/\hbar\omega_X)$ is the 
electron Lorentz factor at which the electron begins to scatter $>1$ X-rays
as it travels along the magnetic loop.
Here $\BQ=m_e^2c^3/e\hbar\approx 4.4\times 10^{13}$~G.
Numerical experiments in BT07 show that the $e^\pm$ discharge
is intermittent on a timescale $\sim r/c$ and 
proceeds in the regime of self-organized criticality. The time-averaged 
current equals the current imposed by the magnetospheric twist, 
$(c/4\pi)\nabla\times\bB$,
and the time-averaged voltage $\Phi_e$ is close to $\Vpm$.

The discharge voltage remains almost independent of the imposed current 
$j$ unless $j$ is reduced below $\js$. The value of $\js$ is unknown but 
small. It may be comparable to $c\rhoGJ$, where $\rhoGJ$ is the corotation 
charge density (Goldreich \& Julian 1969) that should be maintained in the 
magnetosphere in the absence of electric currents. For the purposes of the 
present paper, the following description for $\Phi_e(j)$ will be sufficient,
\be
\label{eq:V}
   \Phi_e(j)=\left\{ \begin{array}{ll}
         \V &  \mbox{if $j\gg\js$} \\
         0 &  \mbox{if $j\ll\js$}
                \end{array}
        \right.
\ee 
where $\js$ is much smaller than the characteristic currents induced by 
the starquake. As will be shown below, neither the value of $\js$ nor the 
behavior of $\Phi_e(j)$ in the transition region $j\sim\js$ matters. 
In essence, $\Phi_e=\V(u)\Theta(j)$, where $\Theta$ is the Heaviside step 
function. In computer simulations, we use a smoothed step function,
\be
\label{eq:W}
  \Phi_e=\V(u)\;W\left(\frac{j}{\js}\right), \qquad
  W(x)=\frac{\arctan(1/\Delta)+\arctan\left[(x-1)/\Delta\right]}
            {\arctan(1/\Delta)+\pi/2},
\ee
where $\Delta\ll 1$. Note that the discharge voltage $\V$ can be 
different for different flux surfaces, i.e. $\V$ in general depends on $u$. 
There is a sharp drop in $\V(u)$ at $u=1$ (voltage is small inside the 
highly conducting star). In computer simulations, we model this drop by 
introducing the factor $\exp\{[\epsilon/(1-u)]^{10}\}$ with $\epsilon\ll 1$. 
The exact form of this factor plays no role for the twist dynamics.

\subsection{Expanding cavity}
\label{sc:front}

Consider again the twist shown in Figure~3, and let us calculate its 
evolution with the new threshold relation between $j$ and $\Phi_e$ 
(eq.~\ref{eq:V}).
The numerical solution of equations~(\ref{eq:evd2}), (\ref{eq:jd}), and 
(\ref{eq:W}) is shown in Figure~5 for the simplest model that assumes 
$\V(u)=const$ in the magnetosphere and $\V(u)=0$ inside the star.

\begin{figure}
\begin{center}
\epsscale{.80}
\plotone{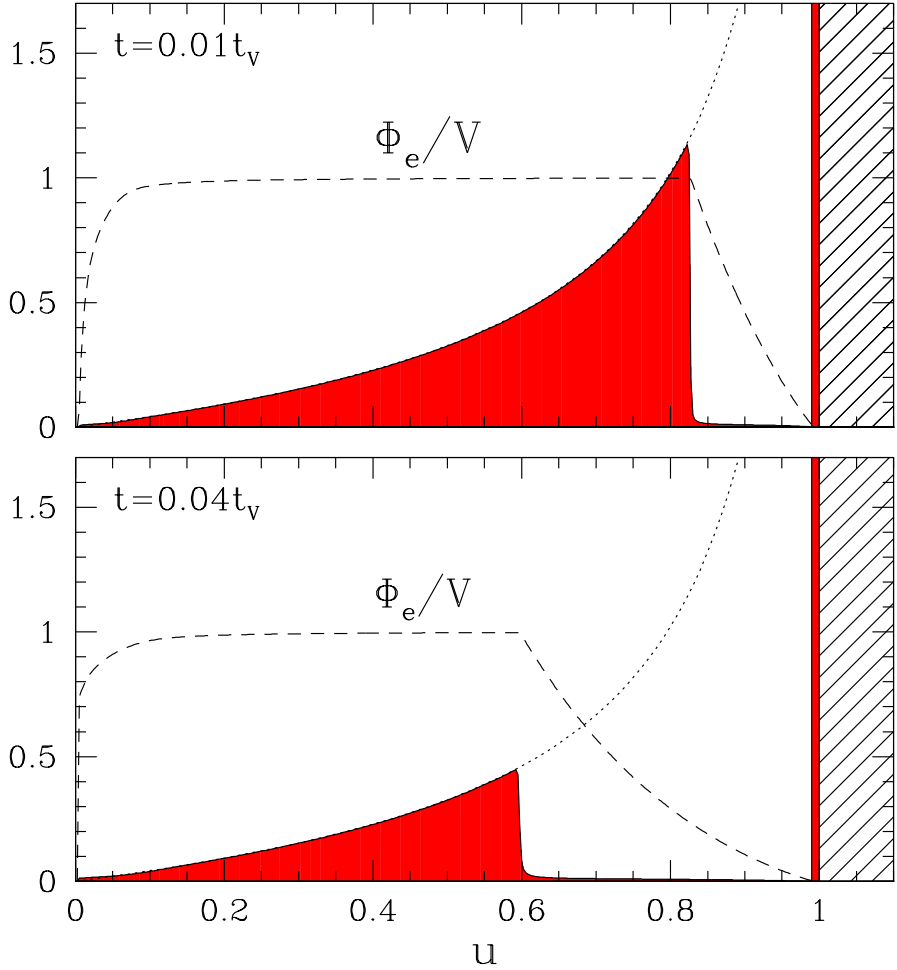}
\caption{
Evolution of the magnetospheric twist with initial uniform amplitude
$\ta_0=0.2$ (Fig.~3). The evolution is caused by the discharge voltage 
$\Phi_e$ that is described by eq.~(\ref{eq:V}), with $\V(u)=const$ at $u<1$ 
and $\V(u)=0$ at $u>1$. {\it Dashed curve} shows $\Phi_e(u)$; it vanishes
in the shaded region $u>1$ (inside the highly conducting star).
{\it Dotted curve} shows the initial current distribution $dI_0/du$. 
{\it Solid curve} shows the current distribution $\partial I/\partial u$ 
at time $t=0.01\tV$ ({\it upper panel}) and $t=0.04\tV$ ({\it lower panel}),
where $\tV=\mu/cR\V$; current is plotted in units of $\Iref=c\mu/4R^2$. 
Immediately following the starquake ($t=0$), a cavity with 
$j\sim j_\star\approx 0$ forms at $u=1$ and grows with time.
Its boundary --- the current front --- moves to smaller $u$ i.e. 
larger $\Rmax=R/u$, erasing the magnetospheric currents.
{\it Shaded area} shows the magnetospheric current $\If$ 
and the current sheet $I_1$ at $u=1$. The total current $\Itot=\If+I_1$
is conserved: the erased magnetospheric current flows in the current sheet. 
}
\end{center}
\end{figure}

Initially, the drop in $\Phi_e$ near $u=1$ is very steep, i.e.
$\partial\Phi_e/\partial u$ is large and negative, and hence the current 
density here is quickly reduced (cf. eq.~\ref{eq:evd1}). The reduction 
continues until $j\sim\js$, which permits $\Phi_e<\V$. Then a smooth profile 
of $0<\Phi_e(u)<\V$ is established in a region $\uf(t)<u<1$. 
The threshold nature of the discharge leads to formation 
of two distinct regions in the magnetosphere: 

\noindent
(1) ``Cavity'' $\uf<u<1$ where $0<\Phi_e<\V$ and $j\sim\js$ (essentially 
zero current). The current originally injected in this region is sucked 
into the star and flows in the current sheet at $u=1$.

\noindent
(2) Region $u<\uf$ where $\Phi_e=\V$. Here the current remains equal to its 
initial value at $t=0$, i.e. the twist remains static.\footnote{
     This is a consequence of $\V(u)=const$. The twist at $u<\uf$ will not 
     remain static if $\V(u)\neq const$ as discussed below.}

\noindent
The voltage profile in the cavity has $\partial\Phi_e/\partial u<0$ 
which implies $\partial I/\partial t<0$ (eq.~\ref{eq:evd2}). Hence the 
cavity must grow, as indeed seen in the simulation.
The structure of the untwisting magnetosphere is shown in Figure~6.

\begin{figure}
\begin{center}
\epsscale{1.0}
\plotone{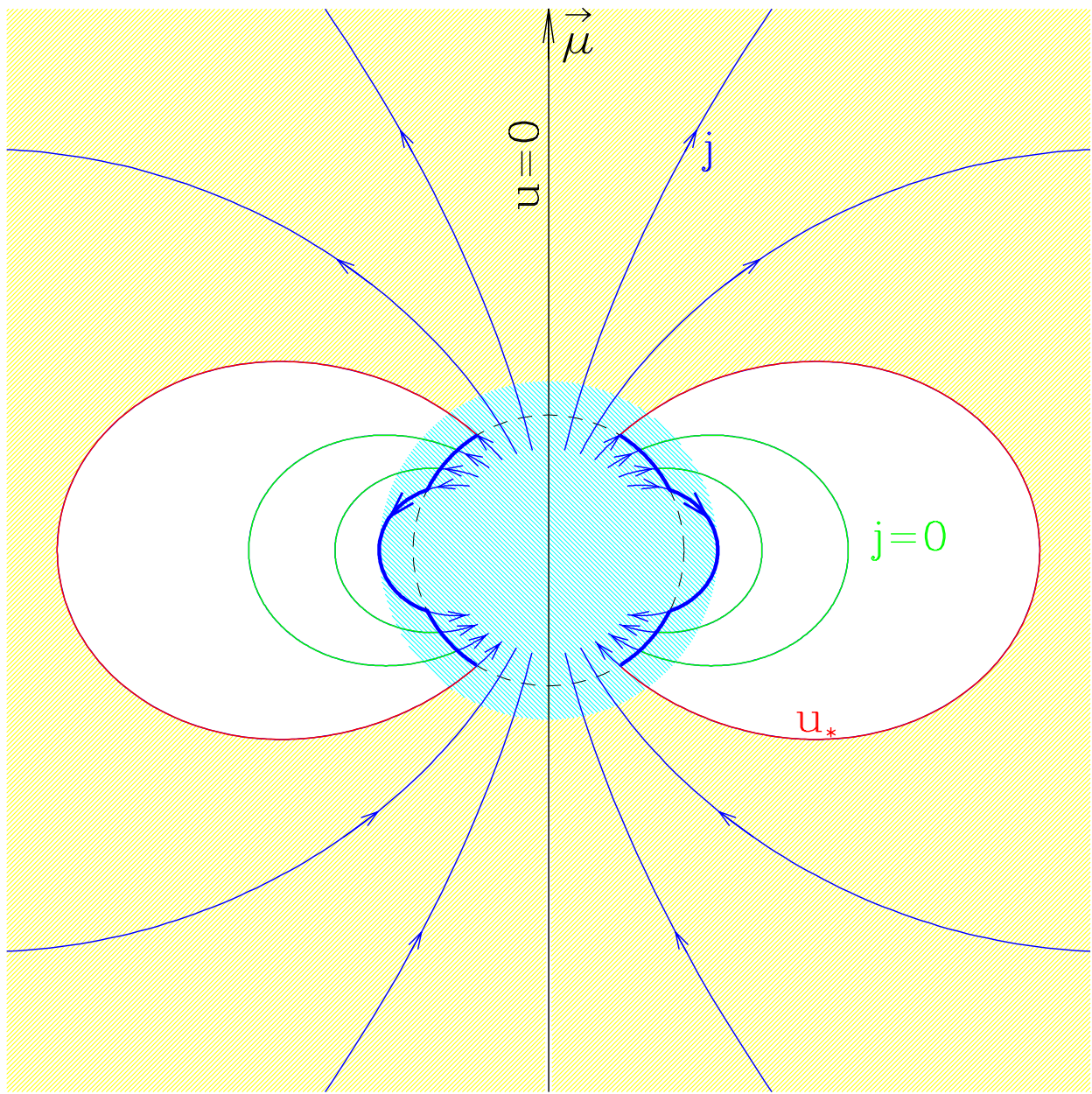}
\caption{
Structure of untwisting dipole magnetosphere (poloidal cross section).
The magnetospheric currents are confined to the j-bundle $u<\uf$  
(the outer region {\it shaded in yellow}). Field lines shown in {\it green} 
are potential ($\nabla\times\bB=0$) and form the inner cavity with 
$j\approx 0$. The cavity is bounded by the current front (located at the 
magnetic flux surface $u=\uf$, shown by {\it red curve}). The front 
expands with time, moving to flux surfaces 
closer to the magnetic axis, and the j-bundle shrinks. The erased currents 
(that initially flowed in the cavity) are closed inside the star and flow in 
the current sheet ({\it thick blue curve}). The final state shown in Fig.~2
is achieved when the current front $\uf$ reaches the magnetic axis $u=0$.
}
\end{center}
\end{figure}

The boundary of the cavity forms a sharp front, which resembles a shock wave.
The front starts as a tiny arc emerging from the star at the magnetic 
equator, and propagates toward smaller $u$ (i.e. outward and poleward, 
to flux surfaces with larger $\Rmax$). We denote its instantaneous position 
by $\uf(t)$.  The profile of the front --- the shape of $j(u)$ near $\uf$ 
--- is controlled by the behavior of $\Phi_e$ at $j\sim \js$.\footnote{
    The profile of the front is controlled by the form of function $W(j/\js)$
    in eq.~(\ref{eq:W}); in the simulation shown in Fig.~5 we chose 
    $\js=10^{-2}cB_0/8\pi R$ and $\Delta=0.2$.} 
However, we are not interested in the exact profile of the front.
It is sufficient to know that it is steep and can be treated as a step
function. The quantity of interest is the {\it speed} of the front 
propagation $d\uf/dt$. In the limit of small $\js$ (steep front), one can use
\be
  \Phi_e=\V(u)\Theta(j),
\ee 
and derive an explicit expression for $d\uf/dt$ (see Appendix B),
\be
\label{eq:front}
\displaystyle
   \frac{d\uf}{dt}=-\frac{\displaystyle \V(\uf)\,\frac{\uf(1+1.4\uf)}{1-\uf}
      +\frac{\uf^2\V^\prime(\uf)}{\sqrt{1-\uf/\uc}}}
     {\displaystyle \frac{4R}{c^2}\left.\frac{dI_0}{du}\right|_{\uf}
  +\left.\frac{d}{du}\frac{u^2\V^\prime(u)}{\sqrt{1-u/\uc}}\right|_{\uf} t},
\ee
where $\V^\prime=d\V/du$ and $I_0(u)\equiv I(u,0)$ is the initial current
function. This ordinary differential equation can be solved for $\uf(t)$.
If $\V^\prime=0$ (as in the model in Fig.~5) the front equation simplifies to
\be
\label{eq:front1}
   \frac{d\uf}{dt}=-\frac{c^2\V\,\uf(1+1.4\uf)}{4R\,(1-\uf)(dI_0/du)_{\uf}}.
\ee
The history of the front propagation is shown in Figure~7.
The front starts with extremely high speed near $u=1$, then decelerates 
and approaches the axis $u=0$ with $d\uf/dt=-cR\V/2\mu\ta_0$.
It reaches $u=0$ (and erases all of the twist) at $\tend=\mu\ta_0/cR\V$.
                                                                                
\begin{figure}
\begin{center}
\epsscale{.9}
\plotone{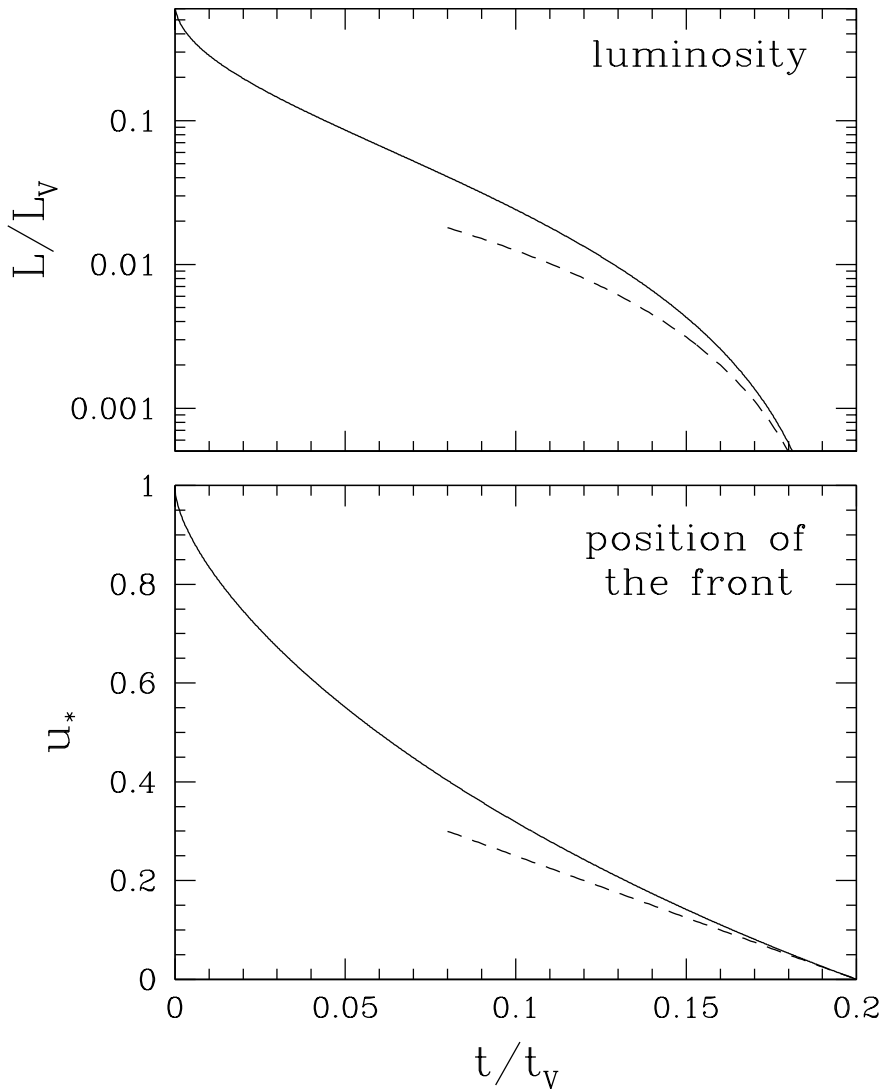}
\caption{
{\it Lower panel}: propagation of the current front $\uf(t)$ in the 
magnetosphere with the initial twist $\ta_0(u)=0.2$ and a discharge 
voltage $\V(u)=const$. Time is expressed in units of $\tV=\mu/cR\V$.
Immediately after the starquake, the front emerges from the star at the 
magnetic equator and begins to expand outward and poleward. The front reaches 
$u=0$ (the magnetic axis) and erases all of the magnetospheric current 
in a finite time $\tend=\ta_0\tV$. The dashed line shows the slope of 
$\uf(t)$ when $\uf$ approaches 0. This slope equals $-(2\ta_0)^{-1}$.
{\it Upper panel}: evolution of the dissipation power $L$ (magnetospheric 
luminosity). $L$ is expressed in units of $\LV=\V\Iref=c\mu\V/4R^2$.
The dashed curve corresponds to the dashed line in the bottom panel.
Both panels show the model with $\ta_0=0.2$.  Similar plots for twists with 
different initial $\ta_0=const$ are obtained by simple stretching of the 
time coordinate $t\rightarrow t(\ta_0/0.2)$; the evolution slows down by 
the factor $\ta_0/0.2$ for stronger twists.
}
\end{center}
\end{figure}

The current function $I$ of the evolving twist is given by 
equation~(\ref{eq:I_app}) in Appendix~B. The corresponding twist amplitude 
is given by
\be
\label{eq:psi}
   \ta(u,t)=\left\{ \begin{array}{ll} 
 \displaystyle\ta_0(u)+\frac{cR}{\mu}\,\V^\prime(u)\,t            & 0<u<\uf \\
 \displaystyle\frac{4R^2}{c\mu u^2}\sqrt{1-\frac{u}{\uc}}\,\If(t) & \uf<u<1 \\
                  \end{array}
          \right.
\ee
where $\If(t)\equiv I[\uf(t),t]$ is the net current that flows through the 
magnetosphere at time $t$. Equation~(\ref{eq:psi}) together with $\uf(t)$ 
gives a complete analytical description for untwisting magnetospheres
(in the linear-twist approximation, see \S~4.2).

Note that the numerical simulation shown in Figure~5 gives $\Phi_e/\V<1$ 
for $u\ll 1$, i.e. for flux surfaces with footpoints near the magnetic axis.
This is not predicted by the analytical model, which assumes $\js=0$. 
The drop appears in the numerical simulation at a finite $u\ll 1$ because 
the numerical model assumes a finite $\js$. The twist with $\ta_0(u)=const$ 
has the current density near the axis $j(u)\propto u^2$ and hence, 
in a small polar region where $j\simlt\js$, $\Phi_e$ must drop. 
In the limit $\js\rightarrow 0$, this effect disappears in the sense that
the polar region with $j\sim\js$ shrinks to one point $u=0$.

The twist behavior on the axis becomes important when the spindown of 
the star is of interest. The model in Figure~5 assumes a finite $\js$, 
however, it does not take into account rotation of the star; therefore
$j(0)=0$ and $\Phi_e(0)=0$. Rotation with angular velocity $\Omega$ implies 
the additional twisting and opening of the field lines that extend to the 
light cylinder $\Rlc=c/\Omega$. This persistent ``external'' twisting 
induces small but finite currents on the magnetic dipole axis, along the 
bundle of open field lines. The open bundle has the parameter
$\upc=R/\Rlc\ll 1$. In particular, magnetars have $\upc\simlt 10^{-4}$.
As long as the behavior of the closed magnetosphere $u>\upc$ is concerned, 
the rotation can be neglected, and $\js\rightarrow 0$ is a good approximation.


\subsection{j-bundle with growing twist}
\label{sc:growth}

As the cavity expands, the current-carrying region $u<\uf$ shrinks.
We will call the current-carrying bundle of magnetic field lines 
``j-bundle,'' for brevity. The numerical model of \S~\ref{sc:front} 
(Figs.~5 and 7) assumed that the discharge voltage is the same for all 
magnetospheric flux surfaces, $\V(u)=const$. Then the twist remains static 
inside the j-bundle. It freezes and waits while it is eaten by the 
expanding front $\uf(t)$. 

In contrast, if $\V(u)\neq const$, the twist in the j-bundle will change 
linearly with time as it waits for the front to come. This change is 
described by equation~(\ref{eq:evd1}) [note that $\Phi_e=\V(u)$ inside 
the bundle $u<\uf$] or equation~(\ref{eq:psi}). The twist amplitude $\ta$ 
decreases if $\V^\prime<0$ and grows if $\V^\prime>0$. Observational data 
(discussed below) suggest $\V^\prime>0$ and the growth of $\ta$ near the 
axis $u=0$. Despite the twist growth at small $u$, its total energy 
$\Etor$ is decreasing with time as the j-bundle shrinks. This evolution 
is consistent with the energy conservation law (eq.~\ref{eq:en}). 

For illustration, we calculated the same model as in Figure~5 but with 
new threshold voltage $\V(u)=(0.04+2u)^{1/2}\bar{\V}$, where $\bar{\V}$
approximately equals the average of $\V(u)$ in the magnetosphere $0<u<1$.
Figure~8 shows the evolution of the twist amplitude $\ta$ in this case
and compares it with the case of $\V(u)=const$. 

\begin{figure}
\begin{center}
\epsscale{.90}
\plotone{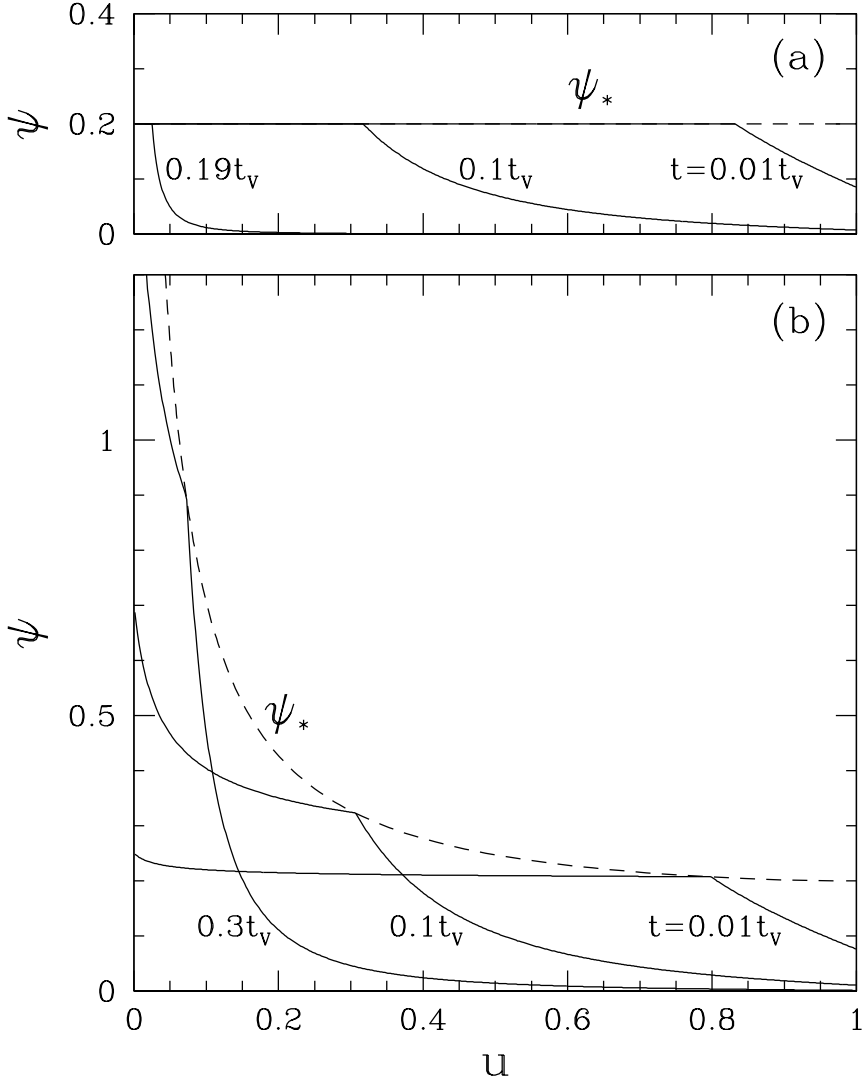}
\caption{
Evolution of the twist profile $\ta(u)$.
{\it (a)} Model with $\V(u)=const$ (same model as in Fig.~5). 
{\it (b)} Similar model but with $\V(u)=(0.02+2u)^{1/2}\bar{\V}$.
{\it Solid curves} show $\ta(u,t)$ at three different moments of time $t$.
Time is expressed in units of $\tV=\mu/cR\V$ in panel (a) and 
$\tV=\mu/cR\bar{\V}$ in panel (b). 
{\it Dashed curve} shows $\ta_\star=\ta(\uf)$ --- the twist amplitude at 
the boundary of the cavity throughout the entire history of its 
expansion from $\uf=1$ to $\uf=0$.
}
\end{center}
\end{figure}

The twist evolution is described by simple analytical formulas when the 
j-bundle is narrow, $\uf\ll 1$. In the leading order of $\uf\ll 1$ the 
front equation becomes
\be
   \frac{d\uf}{dt}=-\frac{\uf \V(\uf)}{(4R/c^2)\,I_0^\prime(\uf)
            +\left.\left[u^2\V^\prime\right]^\prime\right|_{\uf} t}
   \left[1+{\cal O}(\uf)\right],  \qquad \uf\ll 1,
\ee
where prime denotes the differentiation with respect to $u$.
The relation between $I$ and $\ta$ (eq.~\ref{eq:ta2}) gives
$I_0(u)=(c\mu\ta_0/4R^2)u^2+{\cal O}(u^3)$ where $\ta_0(u)=\ta(u,0)$ 
is the initial amplitude of the twist. Then we get, 
\be
   \frac{d\uf}{dt}=-\frac{\V(0)}
      {[2\mu\ta_0(0)/cR]+2\V^\prime(0)\,t}\,\left[1+{\cal O}(\uf)\right].
\ee
If $\V^\prime(0)\neq 0$, integration of this equation yields 
\be
\label{eq:uf_late}
   \uf(t)=\frac{\V(0)}{2\V^\prime(0)}\,\ln\frac{t_0+\tend}{t_0+t}, 
   \qquad t_0\equiv \frac{\mu\ta_0(0)}{cR\V^\prime(0)},
\ee
where $\tend$ is the time when the front reaches $u=0$; it is finite in 
all cases. 

The twist amplitude $\ta(u,t)$ inside the j-bundle grows according to
equation~(\ref{eq:psi}) (unless $\ta$ reaches $\tamax\simgt 1$),
\be
\label{eq:ta_late}
   \ta(0,t)=\ta(0,0)\left(1+\frac{t}{t_0}\right), \qquad 0<t<\tend.
\ee
It grows by a large factor if $\tend\gg t_0$. 

For example, consider a model with linear 
$\V(u)=\V(0)+\V^\prime(0)u$ at $u<\hat{u}$ and constant 
$\V(u)=\V(\hat{u})$ at $u>\hat{u}$. Suppose $\V(\hat{u})\gg \V(0)$. 
Then we find,
\be
\label{eq:ta_toy}
   \frac{\ta(0,\tend)}{\ta(0,0)}
   \sim \exp\left[2\left(\frac{\V(\hat{u})}{\V(0)}-1\right)\right]\gg 1.
\ee
The drop in $\V(\uf)$ at small $\uf$ delays the arrival of the front to 
$u=0$ and gives an exponentially longer time for the twist growth at $u=0$. 
Thus, even a small twist with $\ta_0\ll 1$ can grow to $\tamax$ 
inside the j-bundle. Further growth is impeded by the MHD instability.

\subsection{Localized starquakes}
\label{sc:ring}

Starquakes may rotate part of the crust and leave the rest of it untouched. 
Suppose that a ring $u_2<u<u_1$ has been rotated. 
Then the twist $\ta\neq 0$ is created only in the region $u_2<u<u_1$.
This implies $I(u>u_1)=0$, i.e. the net ejected current is zero, and 
hence the current density must change sign at some $\up$ in the region 
$u_2<u<u_1$. The current function $I_0(u)$ reaches a maximum at $\up$.
An example of such a localized twist and its evolution are shown in Figure~9. 
Voltage $\V(u)=(\frac{1}{2}+u)\bar{\V}\neq const$ is assumed in the model.

\begin{figure}
\begin{center}
\epsscale{.80}
\plotone{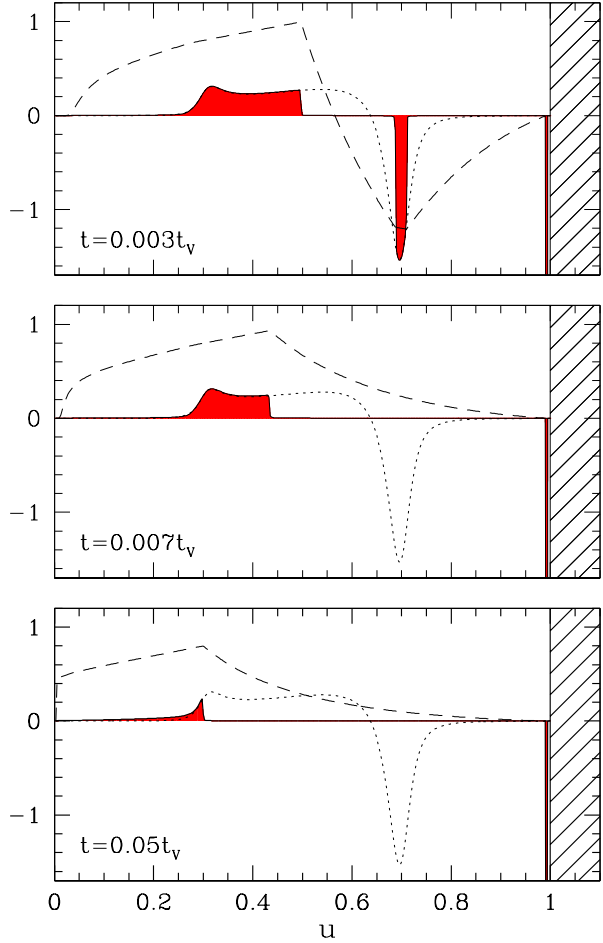}
\caption{
Evolution of the magnetospheric twist that is created by rotation of 
a crustal ring. The initial twist amplitude $\ta_0\approx 0.2$ in the
region $0.3<u<0.7$ and close to zero outside this region. The corresponding
initial current distribution $dI_0/du$ is shown by {\it dotted curve}.
Current is plotted in units of $\Iref=c\mu/4R^2$.
The discharge voltage $\V(u)=(\frac{1}{2}+u)\bar{\V}$ is assumed in this model
[$\bar{\V}$ is the average of $\V(u)$ in the magnetosphere $0<u<1$].
{\it Solid curve} shows the current distribution $\partial I/\partial u$
at time $t=0.003\tV$ ({\it upper panel}), $t=0.007\tV$ ({\it middle panel})
and $t=0.05\tV$ ({\it lower panel}), where $\tV=\mu/cR\bar{\V}$.
}
\end{center}
\end{figure}

\begin{figure}
\begin{center}
\epsscale{1.0}
\plotone{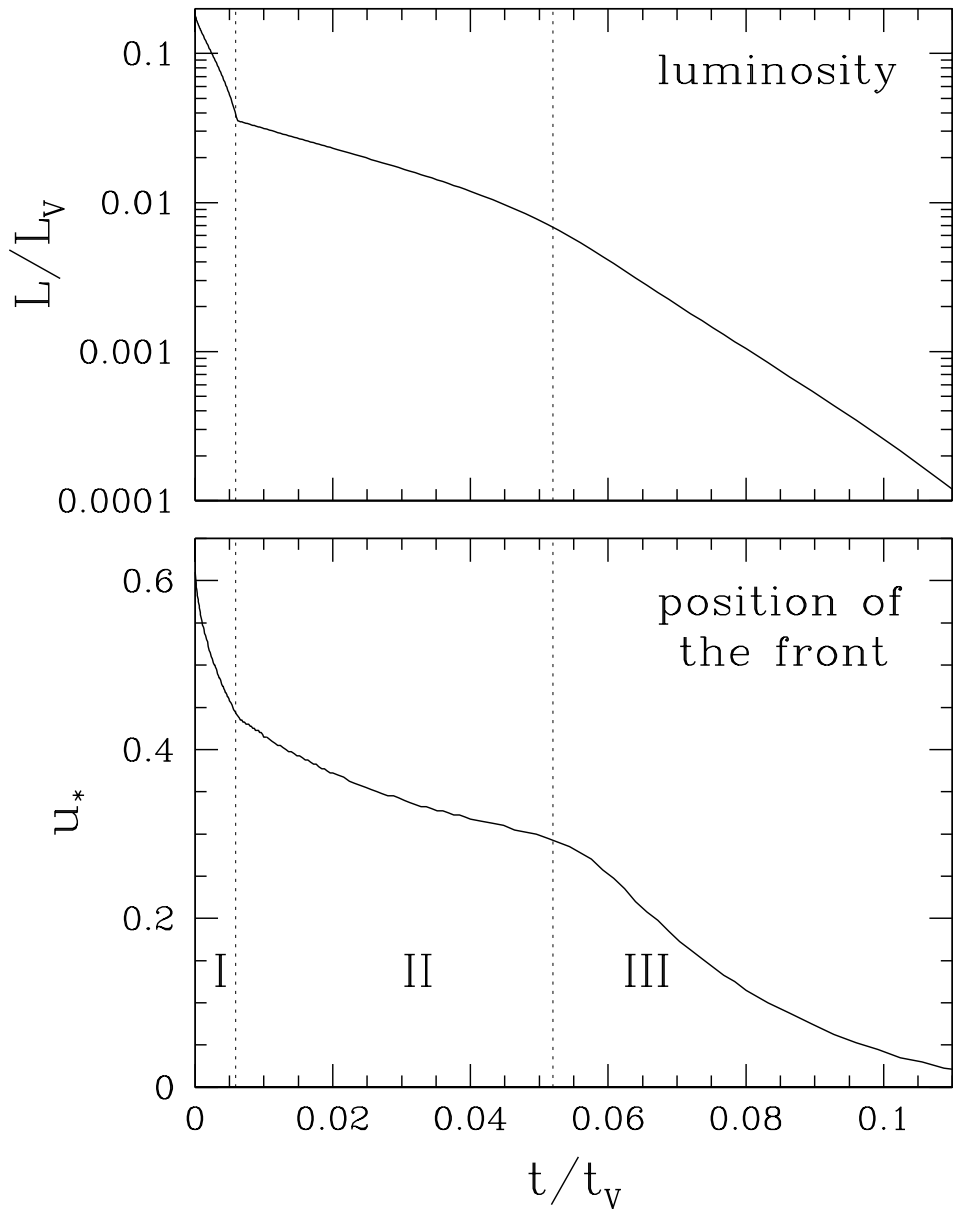}
\caption{
{\it Upper panel}: evolution of the dissipated power $L$ in the model 
shown in Fig.~9. $L$ is expressed in units of 
$\LV=\Iref\bar{\V}=c\mu\bar{\V}/4R^2$.
Time is expressed in units of $\tV=\mu/cR\bar{\V}$.
{\it Lower panel}: outward-propagating front that erases the ejected
positive current (cf. Fig.~9). This front starts at $\up\approx 0.63$ 
where the ejected current density $j\propto dI_0/du$ changes sign. 
Vertical dotted lines mark stages I-III in the twist evolution 
(see the text).
}
\end{center}
\end{figure}

Three stages may be noted in the evolution of the ring twist:

I. One current front is launched from $u=1$. It immediately jumps to the 
twist boundary $u_1$, and continues to propagate toward smaller $u$.
In addition, two divergent current fronts are immediately launched from 
$u=\up$. Thus, two cavities form in the magnetosphere. Both cavities grow 
until they merge: the two fronts erasing the spike of negative current 
(Fig.~9) eventually meet and ``annihilate.'' Stage I ends at this point 
(at time $t\approx 0.005\tV$ where $\tV=\mu/cR\bar{\V}$). 

II. The merged cavity continues to expand poleward and erase the remaining 
spike of positive current at smaller $u$. All of the initially injected 
current is erased (and stage~II ends) at $t\approx 0.05\tV$.

III. The current front proceeds toward the axis, erasing the currents
that have grown there (from zero) since the beginning of the twist evolution.

Voltage $\V(u)\neq const$ was chosen in the model to 
allow the twist growth near the axis. The simulation shows, however, 
that the growth is slow compared to the expansion of the cavity.
Only at the last stage III does the grown current create a significant twist 
$\ta$ near the axis and somewhat decelerate the expansion of the cavity.

Figure~10 shows the evolution of the twist luminosity $L(t)$. Its decrease
is quickest during stage I, as the spike of negative current is quickly 
erased. Figure~10 also shows the propagation of the current front $\uf(t)$ 
that starts at $\up$ and moves to smaller $u$, erasing the ejected positive 
current.

Instead of a ring $u_2<u<u_1$, an axisymmetric localized starquake may
rotate a cap $u<u_1$. The cap-twist and ring-twist evolve in a similar way.
In particular, stages I and II are similar.
However, stage III is absent for the cap-twist. It has $u_2=0$, i.e. the 
initially ejected positive currents occupy the enire region around the 
polar axis. Erasing these currents takes a longer time. As a result, the 
twist grows to a large amplitude at $u\ll 1$ before the cavity reaches 
the axis (\S~\ref{sc:growth}). Then a maximally twisted narrow bundle with 
$\ta=\tamax$ forms.


\section{Observational effects}

\subsection{Luminosity}
\label{sc:bundle}

Energy dissipation in the untwisting magnetosphere is confined to the bundle 
of current-carrying field lines $u<\uf$ (j-bundle). It can be a bright 
source of radiation, with luminosity equal to the rate of Ohmic dissipation 
$L$. The luminosity generally decreases as the magnetosphere untwists. 
Examples of the evolution of $L(t)$ are shown in Figures~7 and 10.

A large fraction of the dissipated power may be radiated quasi-thermally at 
the footprints of the j-bundle as the accelerated magnetospheric particles 
run into the star (BT07), creating a hot spot on the surface 
$\theta<\theta_\star$.  The area of this spot is given by,
\be
\label{eq:area}
   A\approx \pi(R\sin\theta_\star)^2=\pi R^2\uf.
\ee
As the cavity expands in an untwisting magnetosphere (Fig.~6), the spot 
shrinks.

The evolution of a narrow j-bundle ($\uf\ll 1$) with a uniform twist $\ta$ 
(e.g. $\ta\approx\tamax\sim 1$) is described by simple formulas.
The free energy of the twist (eq.~\ref{eq:en0}) is then given 
by\footnote{Only 1/4 of 
    $\Etor$ resides in the region 
    $u<\uf$; 3/4 of the twist energy is contained in the potential region 
    $u>\uf$. The fact that $\nabla\times\bB=0$ in the potential region does 
    not imply that $B_\phi=0$; $B_\phi$ is determined by eq.~(\ref{eq:Bphi}).
}
\be
\label{eq:energy}
  \Etor=\frac{\mu}{2cR}\int_0^1 I(u)\,\ta(u)\,du
       \approx \frac{\mu^2\ta^2\uf^3}{6R^3}
       \approx 4\times 10^{44}\,B_{14}^2\,R_6^3\,\ta^2\,\uf^3\;{\rm erg},
\ee
where $R_6\equiv R/10^6$~cm, $B_{14}\equiv \Bpole/10^{14}$~G, and
$\Bpole\equiv 2\mu/R^3$. The luminosity of the j-bundle with 
$\ta\approx const$ can be immediately calculated for a given voltage $\V(u)$, 
\be
\label{eq:lum0}
  L=\int_0^{\If}\V\,dI
    =\frac{c\mu\psi}{4R^2}\int_0^{\uf} \V(u)\,
                         d\left(\frac{u^2}{\sqrt{1-u/\uc}}\right).
\ee
The simplest model with $\V(u)=const$ gives
\be
\label{eq:lum}
   L=\V\If\approx \frac{c\mu}{4R^2}\,\ta\,\V\,\uf^2
   \approx 1.3\times 10^{36}\,B_{14}\,R_6\,\ta\,\V_9\,\uf^2\;{\rm erg~s}^{-1}.
\ee
where $\V_9\equiv\V/10^9$~V. The j-bundle with $\V(u)=const$ shrinks 
according to equation~(\ref{eq:front1}), which yields at $\uf\ll 1$
\be
\label{eq:spot}
     \frac{d\uf}{dt}\approx -\frac{cR\V}{2\mu\ta} \qquad
    \Rightarrow \quad \uf(t)\approx \frac{cR\V}{2\mu\ta}\left(\tend-t\right).
\ee
This equation is also easy to derive from $d\Etor/dt=-L$, using 
equations~(\ref{eq:energy}) and (\ref{eq:lum}).
The evolution timescale of the luminosity is given by
\be
\label{eq:tev}
  \tev=-\frac{L}{dL/dt}\approx \frac{\mu\uf}{cR\V}
      \approx 15\,\V_9^{-1}\,B_{14}\,R_6^2\,\ta\,\uf\;{\rm yr}.
\ee

Approximate formulas can also be derived for $\V(u)\neq const$.
For example, for $\V(u)=\V(0)+\V^\prime(0)u$ we find
\be
\label{eq:lum2}
   L=\frac{c\mu\V(0)}{4R^2}\,\ta\,\uf^2\left[1+
     \left(\frac{2\V^\prime(0)}{3\V(0)}+\frac{1}{2\uc}\right)\uf
        +{\cal O}\left(\uf^2\right)\right].
\ee
This equation again assumes $\ta(u<\uf)\approx const$. It may approximately
describe e.g. the maximally twisted j-bundle, $\ta\approx\tamax$.

The decay of the twist luminosity $L(t)$ was previously estimated assuming 
that the current is decaying uniformly in the twisted region, which gave a 
linear $L(t)\propto t-\tend$ (BT07).
The electrodynamic theory developed in this paper shows that the 
untwisting is strongly {\it non-uniform}: the twist is erased by 
the propagating front that resembles a shock wave.
The speed of this front depends on the initial twist configuration $\ta_0(u)$.
Similar to the simple estimate $L(t)\propto t-\tend$, we find that the 
twist is erased in a finite time and $L(t)$ vanishes at $\tend$ 
(unless new starquakes occur).
However, no universal linear shape of $L(t)$ is predicted.
In some cases $L(t)$ may resemble a linear decay (e.g segment II in 
Fig.~10 is almost linear in a linear plot).
In observed sources, $L(t)$ was close to linear in 
AXP~1E~1048.1-5937 (Dib, Kaspi, \& Gavriil 2008) and non-linear 
in \XTE (Gotthelf \& Halpern 2007).

\subsection{Nonthermal radiation}
\label{sc:nonth}

The energy released in the j-bundle can power nonthermal magnetospheric 
emission. Such emission is observed in most magnetars.
Two distinct non-thermal components are detected in their spectra:
(1) soft X-ray tail that extends from 1~keV to $\sim 10-20$~keV with a 
photon index $\Gamma\sim 2-4$ (e.g. Woods \& Thompson 2006 and refs. therein), 
and (2) hard X-ray component that extends to $\sim 300$~keV with
$\Gamma\sim 0.8-1.5$ (e.g. Kuiper et al. 2008 and refs. therein).

The 1-20~keV tail is usually explained by resonant scattering of thermal 
radiation by the magnetospheric plasma (Thompson et al. 2002; Lyutikov \& 
Gavriil 2006; Fernandez \& Thompson 2007; Nobili, Turolla, \& Zane 2008; 
Rea et al. 2008). Ions resonantly scatter thermal photons near the star 
where $\hbar eB/m_ic\sim$ keV, and $e^\pm$ scatter at radii $r\sim 10R$ 
where $\hbar eB/m_ec\sim$ keV. 
The growth of cavity in the untwisting magnetosphere implies that the 
scattering in the inner magnetosphere is suppressed, because the dense 
plasma is confined to the narrow j-bundle. At larger radii $r\sim R/\uf$, 
the j-bundle broadens and forms an outer corona that subtends a large 
solid angle as viewed from the star. The cyclotron energy in this region is 
\be 
 \hbar\,\frac{eB}{m_ec}\sim 1 \left(\frac{B_{\rm pole}}{10^{14}{\rm~G}}\right)
                            \left(\frac{\uf}{0.1}\right)^3 {\rm ~keV},
\ee
and resonant scattering by $e^\pm$ can give 1-20~keV photons. 
The luminosity expected from a strongly twisted j-bundle 
with $\uf\sim 0.1$ is consistent with the 
typical nonthermal luminosity of magnetars, $L\sim 10^{35}$~erg/s 
(see eq.~\ref{eq:lum} and substitute the typical 
$\mu\sim 3\times 10^{32}$~G~cm$^3$ and $\ta=1-2$).
If no new starquakes occur, the j-bundle must shrink toward the magnetic 
axis with time, and $\uf$ will be reduced below 0.1. Then the resonant 
scattering must be suppressed. This suppression is caused by two reasons: 
$\hbar eB/m_ec$ in the outer corona $r\sim R/\uf$ decreases below
keV (it is proportional to $\uf^3$), and the power dissipated in the 
j-bundle becomes small as $L\propto\uf^2$.

The j-bundle with $\uf\sim 0.1-0.2$ dissipates sufficient energy to explain 
also the hard X-ray component. This component was detected in three 
anomalous X-ray pulsars (AXPs) and two soft gamma-ray repeaters (SGRs) 
(see Kuiper et al. 2008 for a recent review).
In AXPs, the hard X-ray emission has a huge pulsed fraction, approaching 
100\% at high energies (Kuiper et al. 2006; den Hartog et al. 2008a,b).
This may be explained if the emission is produced in the narrow j-bundle
near the star. 

Observations indicate that the nonthermal emission can be stable on 
timescales as long as a decade (den Hartog et al. 2008a). 
There may be two reasons for this stability:
(1) The discharge voltage $\V$ is relatively low (below 1~GeV)
and the j-bundle is relatively thick, $\uf\sim 0.2$. Then the 
untwisting timescale becomes long (see eq.~\ref{eq:tev}). 
(2) The j-bundle is kept in a quasi-steady, maximally twisted state 
$\ta\sim \tamax$ by frequently repeating (possibly continual) shearing 
motion of the crust in a fixed region $u<\uf$.

The plasma filling the j-bundle may also produce optical and infrared 
radiation by mechanisms discussed in BT07. Besides, it can be a bright 
source of radio waves, as suggested in section \S~\ref{sc:radio}.

\subsection{Outer magnetosphere and spindown rate of the star}
\label{sc:torque}

The magnetospheric twist is expected to impact the spindown rate when the 
twist amplitude is large, $\ta\simgt 1$. This impact may occur in two ways.

(1) The strong twist inflates the poloidal field lines and increases the 
magnetic field at the light cylinder, which leads to stronger spindown 
torque acting on the star (Thompson et al. 2002). The poloidal inflation is 
common for twisted configurations. It is seen e.g. in the self-similarly 
twisted dipole (Wolfson 1995). A similar inflation must occur when the 
currents are confined to the j-bundle. Its calculation will require a 
full nonlinear model. Here we limit our consideration to simple estimates. 

One can think of poloidal inflation as an increase of magnetic dipole moment 
with radius. This effect is easiest to evaluate for moderate twists
$\ta<1$. The current $dI$ flowing along a bundle of twisted field lines $du$ 
creates a toroidal current $dI_\phi\approx dI\,\ta(u)/2\pi$. The field lines 
carrying this current extend to $\Rmax=R/u$, and the dipole moment created 
by $dI_\phi$ is $d\mu\sim dI_\phi \Rmax^2/c$. Integrating over $u$, we 
obtain the net change of dipole moment of the star due to the twist 
\be
\label{eq:dmu1}
  \Delta\mu\sim \frac{R^2}{2\pi c}\int \frac{\ta(u)}{u^2}\,dI. 
\ee
The j-bundle $\upc<u<\uf$ with a uniform twist $\ta$ 
increases the dipole moment of the star by
\be
\label{eq:dmu2}
  \frac{\Delta\mu}{\mu}\sim \frac{\ta^2}{4\pi}\,\ln\frac{\uf}{\upc}.
\ee
This effect is quadratic in $\ta$ and quickly becomes small for $\ta<1$.
For strong twists $\ta\sim\tamax$, the estimate (\ref{eq:dmu2}) must be 
replaced by a full nonlinear calculation. Qualitatively, it suggests that 
$\Delta\mu/\mu$ is reduced as the j-bundle shrinks ($\uf$ decreases). 
Therefore, the spindown torque is expected to be reduced with time and 
gradually come back to the standard dipole torque as $\uf\rightarrow\upc$.

(2) When the twist has grown to $\tamax$, it will begin to ``boil over'' 
through a repeated instability. 
The energy that would be stored in the magnetosphere if $\ta$ kept growing 
above $\tamax$ is then carried away by an intermittent magnetic outflow.
The outflow may also carry away a significant angular momentum.

The outflow can be generated where the overtwisted field lines open up. 
This opening occurs because the twist is constantly pumped near the axis by 
$\V^\prime>0$. A possible structure of untwisting magnetospheres with 
$\V^\prime>0$ is schematically shown in Figure~11. It resembles
the picture of a rotationally powered pulsar (see e.g. Arons 2008 for a 
review), however the opening is caused by the internal resistive 
dynamics of the magnetosphere and depends on the profile of $\V(u)$. 
By contrast, in ordinary pulsars the twist is pumped by the star rotation. 

\begin{figure}
\begin{center}
\epsscale{1.0}
\plotone{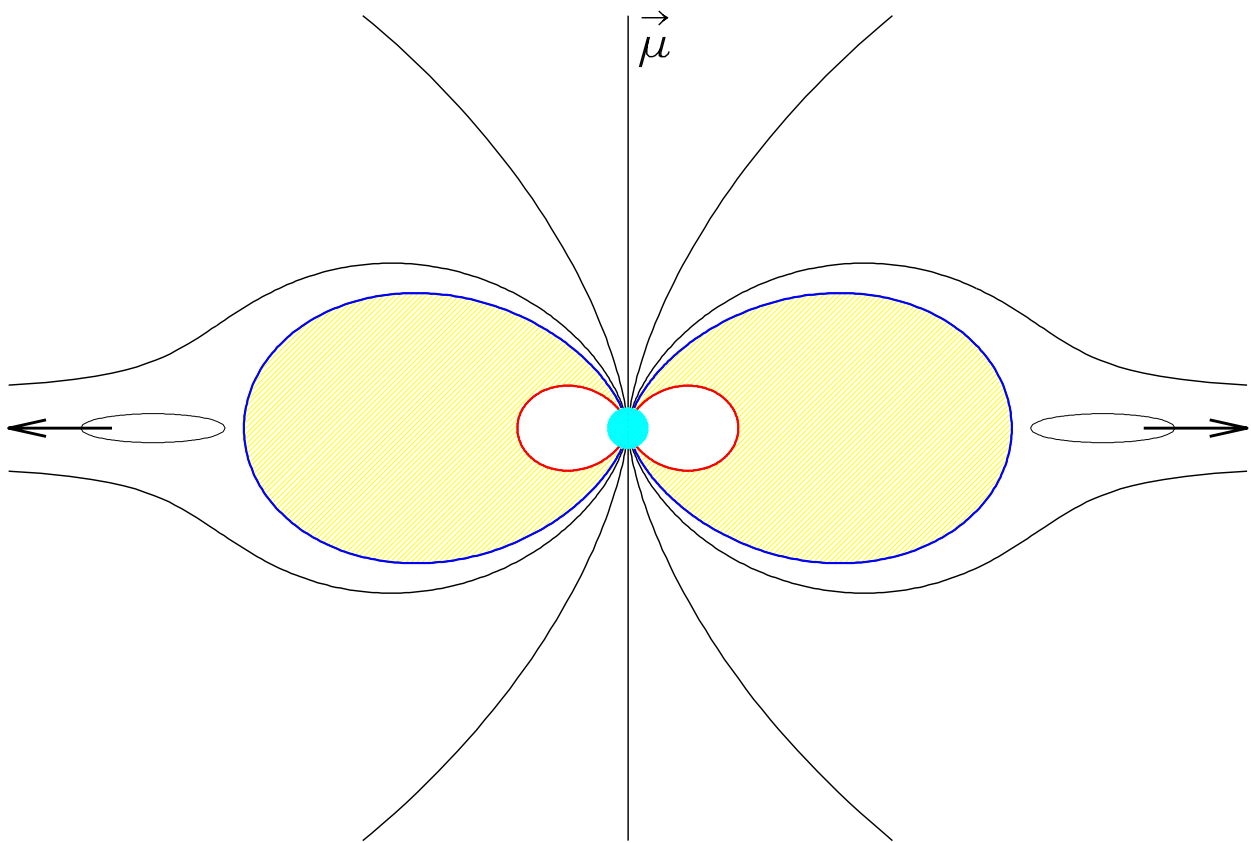}
\caption{Possible structure of untwisting magnetospheres (poloidal cross
section). The growing twist near the axis inflates the outer magnetosphere 
until its field lines open. Then the j-bundle ({\it shaded in yellow}) 
becomes confined between the last closed flux surface ({\it blue curve}) 
and the inner cavity ({\it red curve}), which expands with time. 
An equatorial outflow is expected to form 
just outside the last closed flux surface. The outflow is driven by 
$d\V/du>0$, which forces the twist to grow until part of the overtwisted 
field lines reconnect away from the star. The small circle in the center
(shaded in {\it cyan}) shows the neutron star.
}
\end{center}
\end{figure}

Mechanisms (1) and (2) start immediately after the starquake if it implants 
a strong initial twist $\ta_0>1$ near the magnetic axis. If $\ta_0<1$, 
the spindown torque may not be affected until $\ta$ grows to $\sim 1$, 
which takes time (cf. eq.~\ref{eq:evd1} or eq.~\ref{eq:psi})
\be
\label{eq:tg}
    \tg\approx \frac{\mu}{cR\V^\prime},
\ee
where $\V^\prime=d\V/du$ is evaluated near the axis $u\approx 0$;
$\V^\prime$ may be large, because $u$ is small. For example, a change in
$\V$ from $\V=10^9$~V at $u=0$ to $2\times 10^9$~V at $u=10^{-2}$ 
corresponds to $\V^\prime=10^{11}$~V.
The delay is observed in some objects, with a characteristic $\tg\sim 10^7$~s
(e.g. Gavriil \& Kaspi 2004). This requires $\V^\prime\sim 10^{11}$~V. 

In contrast to luminosity $L(t)$ (which always tends to decrease after 
the starquake), the torque behavior is generally non-monotonic.
If $\ta_0<1$, the torque is expected to grow as $\ta$ grows in the shrinking
j-bundle. (Such an anti-correlation between the torque and the X-ray 
luminosity was observed in 1E~1048.1-5937, see Gavriil \& Kaspi 2004). 
Once $\ta$ has reached $\tamax\sim 1$ the torque should start to decrease,
because the shrinking of the j-bundle at constant $\ta\approx\tamax$ leads 
to the reduction of $\Delta\mu$. The power of the outflow from the 
magnetosphere is also expected to decrease. The simultaneous decrease in 
torque and luminosity was observed, e.g., in \XTE~ (Camilo et al. 2007).


\section{Untwisting magnetosphere in XTE~J1810-197}
\label{sc:XTE}

XTE~J1810-197 is an anomalous X-ray pulsar with period $P=5.54$~s
and estimated dipole magnetic moment $\mu\sim 1.5\times 10^{32}$~G~cm$^3$,
which corresponds to the surface field at the polar cap
$B\sim 3\times 10^{14}$~G (Gotthelf \& Halpern 2007). 
An X-ray outburst was detected from this object in January 2003
(Ibrahim et al. 2004).
Its luminosity approximately followed an exponential decay on
a timescale of 233 days for 3 years (Gotthelf \& Halpern 2007).
During the X-ray decay, the source became radio-bright (Halpern et al. 2005)
and switched on as a powerful radio pulsar with unusual spectrum
and pulse-profile variations (Camilo et al. 2007).
The spindown rate of the star dramatically increased following the outburst.
In the subsequent years, the object gradually evolved toward its quiescent 
(pre-outburst) state.
                                                                                
These observations clearly indicate that the magnetosphere of \XTE~ changed
in the outburst, i.e. the footpoints of field lines must have moved,
imparting a twist to the magnetosphere.  The observational data give 
significant hints about the twist geometry and evolution:
                                                                                
(1) The change in spindown rate suggests that the open field-line bundle 
was affected by a strong twist $\ta\simgt 1$ near the magnetic dipole axis.
On the other hand, $\ta$ is limited to $\tamax={\cal O}(1)$ by the MHD 
instability. Therefore, we infer $\ta={\cal O}(1)$ near the axis.
                                                                                
(2) Already one year after the outburst, the object luminosity was below 
$10^{35}$~erg/s.  The theoretically expected luminosity from a global twist 
with $\ta\sim 1$ (eq.~\ref{eq:lum}) would be much higher: 
$L\sim 3\times 10^{36}\V_9$~erg/s, and it would decay much slower
than observed (eq.~\ref{eq:tev}). Therefore, we conclude that the 
twisted region was small: the current-carrying field lines formed a narrow 
bundle emerging from a small spot on the star surface.
                                                                                
(3) Remarkably, the spot was discovered:
a hot blackbody component with a small emission area was
found in the X-ray spectrum following the outburst (Gotthelf \& Halpern 2007;
Perna \& Gotthelf 2008). Its emission area shrank with time until
the spot became barely detectable (Fig.~12).
                                                                                
(4) The X-ray and radio pulse profiles had almost simultaneous peaks,
consistent with the X-ray hot spot being near the magnetic dipole axis
(Camilo et al. 2007).

\begin{figure}
\begin{center}
\epsscale{.90}
\plotone{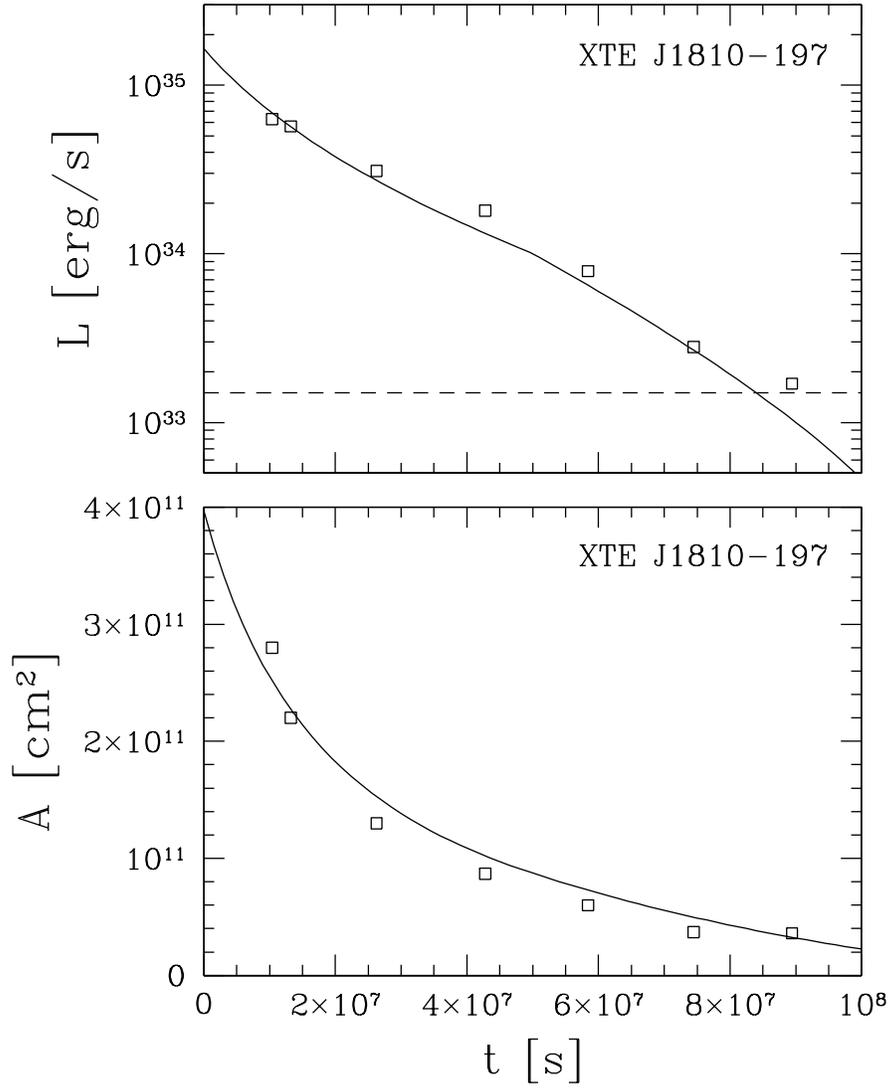}
\caption{
Comparison of the model with the observed evolution of the area $A$ and 
luminosity $L$ of the hot spot formed after the outburst in \XTE.
The data ({\it open squares}) are from Gotthelf \& Halpern (2007).
{\it Dashed line} shows the object luminosity in quiescence.
{\it Solid curve} shows the theoretical model (see the text after 
eq.~\ref{eq:linear}).
}
\end{center}
\end{figure}

These observations are consistent with the theory developed in the present 
paper. The hot spot is explained as the footprint of the j-bundle on the star
(\S~\ref{sc:bundle}). It has a sharp boundary and is shrinking with time as 
the j-bundle shrinks toward the magnetic dipole axis. The large amplitude 
of the twist $\ta\sim 1$ may have been created initially by the starquake, 
but not necessarily: $\ta$ can naturally grow to $\simgt 1$ following the 
starquake (\S~\ref{sc:growth}). A quantitative comparison of the model 
with the observational data is given below.

\subsection{X-ray emitting spot}

Useful preliminary estimates can be made if we assume a uniform voltage 
$\V(u)\approx const$ and uniform twist $\ta(u)\approx const$ across 
the j-bundle $u<\uf$ (\S~\ref{sc:bundle}). Then the produced luminosity 
$L(t)$ is given by equation~(\ref{eq:lum}). Suppose a large fraction of $L$ 
is emitted thermally at the footprint of the j-bundle.
The estimates for $L$ (eq.~\ref{eq:lum}) and its evolution timescale $\tev$
(eq.~\ref{eq:tev}) for a given spot area $A$ (eq.~\ref{eq:area}) may be 
compared with observations. For example, in the fall of 2004, 
the spot area was $A\approx 10^{11}$~cm (and hence $\uf\approx 0.03$).
One then finds that a strongly twisted j-bundle ($\ta\sim 1-1.5$) with 
voltage $\V_9\sim 3$ explains {\it both} observed 
$L\approx 2\times 10^{34}$~erg/s and $\tev\approx 0.6$~yr.\footnote{We use 
here $B_{14}\approx 3$ (Gotthelf \& Halpern 2007).}
                                                                                
The model $\V(u)\approx const$ gives good estimates for $L$, $A$, 
and $\tev$, however, it has a drawback: it is unable to describe the 
possible growth of $\ta$ from a smaller $\ta_0$ before the j-bundle became 
maximally twisted. The growth occurs if $\V^\prime>0$ (\S~\ref{sc:growth}). 
Therefore, we adopt a slightly more general model that includes the next 
(linear) term in the expansion of $\V(u)$ near $u=0$,
\be
\label{eq:linear}
    \V(u)\approx \V_0+\V^\prime\,u, \qquad u\ll 1.
\ee
$\V^\prime$ is unknown and probably large near the axis (see the text after
eq.~\ref{eq:tg}).

A simplest model of a twisted magnetosphere has four parameters: 
magnetic dipole moment of the star $\mu$, radius of the star $R$, the 
initial size of the j-bundle created by the starquake $u_0=\sin^2\theta_0$, 
and the initial amplitude of the twist $\ta_0$ [it equals the angular 
displacement of the crustal cap rotated by the starquake; in our fiducial 
model, the starquake imparts a uniform twist $\psi_0(u<u_0)=const$]. 
The twist evolution after the starquake is controlled by two parameters 
$\V_0$ and $\V^\prime$ that specify voltage $\V(u)$ in 
equation~(\ref{eq:linear}). The evolution is described by 
equations~(\ref{eq:psi}) and (\ref{eq:front}) that we solve numerically. 
If $\ta$ reaches $\tamax={\cal O}(1)$, it is assumed to 
stall at $\tamax$.\footnote{In the simulations, we set $\tamax=1.5$.
    The exact value of the stability threshold needs to be calculated in 
    the full nonlinear twist model.} 
The luminosity of a maximally twisted j-bundle, $L(\uf)$, is 
approximately given by equation~(\ref{eq:lum2}).

Figure~12 compares the model with observations of \XTE.
The star is assumed to have $\mu=1.5\times 10^{32}$~G~cm$^3$ and $R=9$~km.
The parameters of the starquake are $u_0=0.15$ and $\ta_0=0.5$.
The discharge voltage is assumed to drop linearly from $5.5$~GeV at 
$u=0.15$ to 1~GeV at $u=0$ (which corresponds to $\V_0=10^9$~V and 
$\V^\prime=3\times 10^{10}$~V). Although we did not attempt a formal 
fitting of the data, the figure suggests that the model is successful
in explaining the evolution of $L(t)$ and $A(t)$.
The data firmly constrain the voltage to be in 1-6~GeV range, which
is close to the theoretical estimate (eq.~\ref{eq:Vepm}). 

The accuracy of the model is limited by several idealizing assumptions:
uniform initial twist in the starquake region, the linear form of $\V(u)$, 
and axial symmetry.  The actual value of $\mu$ is probably smaller than 
inferred from spindown measurements (\S~\ref{sc:torque}). The linear 
twist-evolution equation becomes approximate when $\ta\sim 1$. Note also 
that our simplest fiducial model assumes that all energy dissipated in 
the j-bundle is emitted thermally at one (e.g. anode) footprint. 
A more realistic model can predict a more complicated spectrum with both 
footprints emitting. Finally, the hot-spot emission may be significantly 
anisotropic (Perna \& Gotthelf 2008), which may increase or reduce its 
apparent luminosity depending of the average inclination of the spot to 
the line of sight.

Note that the observed spectrum was well fitted by a two-temperature
blackbody. The second blackbody component was cooler and had emission 
area comparable to that of the star. No non-thermal component from 
magnetospheric scattering was required by the data. The weakness of the 
scattered component may be explained by the small size $\uf$ of the j-bundle:
the star's radiation does not scatter in the cavity around the j-bundle,
and the probability of scattering in the narrow bundle is low 
(\S~\ref{sc:nonth}).

\subsection{Radio pulsations}
\label{sc:radio}

In ordinary radio pulsars, radio emission is believed to be produced by 
the open field lines passing through the light cylinder, because they are 
the only part of the magnetosphere that carries electric currents. 
In magnetars, these field lines have a tiny $\upc=R/\Rlc\simlt 10^{-4}$ 
where $\Rlc=c/\Omega$ is the light-cylinder radius for a star rotating with 
angular velocity $\Omega$. Radio emission from the open bundles of magnetars 
may be undetectable for two reasons:
(1) $e^\pm$ discharge has a low threshold in magnetars, $\V\sim 10^9$~V.
When this voltage is multiplied with the current in the bundle,
$\Iopen\approx \Iref\upc^2=c\mu/4\Rlc^2$, one obtains the dissipated
power $\Lopen=\V\Iopen\sim 10^{28}\V_9$~erg/s. This power appears to be too
small to feed the observed radio luminosity of \XTE, 
$L_{\rm radio}\sim 10^{30}$~erg/s.
(2) The radio beam may be narrow (because of small $\upc$). Then 
the probability of its passing through our line of sight is small. 
This suggests that radio pulsations are hardly detectable when the magnetar 
is in quiescence, i.e. when its magnetosphere is untwisted.

In contrast, after the starquake, the j-bundle forms. It is much thicker 
and more energetic than the bundle passing through the light cylinder 
and can produce much brighter radio emission with a much broader pulse. 
The untwisting magnetosphere in \XTE~ had $\uf/\upc\simgt 3\times 10^2$.
The net current flowing in the j-bundle is $\sim 10^5$ times larger than 
$\Iopen$, 
\be
 \frac{\If}{\Iopen}\approx \left(\frac{\uf}{\upc}\right)^2\sim 10^5.
\ee
This gives $L/\Lopen\sim 10^5$ (assuming a comparable discharge voltage at 
$\uf$ and $\upc$). A small fraction $\epsradio$ of this luminosity may escape 
as radio waves; $\epsradio\sim 10^{-3}$ is consistent with observations.

Radio waves are efficiently absorbed by the magnetospheric plasma.
Only waves produced close to the magnetic axis, $u<\uesc$, are likely to 
escape, while waves emitted at $u>\uesc$ are trapped in the magnetosphere. 
The value of $\uesc$ depends on the frequency of the wave and the state of 
the plasma (particle density and velocity distribution); $\uesc$ might be 
inferred from the observed opening angle of the radio beam. The radio 
luminosity should quickly decrease when the j-bundle shrinks to $\uf<\uesc$.

Radio pulsar \XTE~ is distinguished
from ordinary radio pulsars by its hard spectrum, very strong linear
polarization, and variable pulse profile. If its emission is produced
on the bundle of closed field lines with $u\gg\upc$, it may be expected 
to be different from ordinary radio pulsars. The spectrum of radio waves 
may form as the sum of emissions from different (frequency-dependent) 
$\uesc$ inside the j-bundle. The sporadic changes in the pulse profile may 
be caused by the instabilities in the maximally-twisted outer magnetosphere
(\S~\ref{sc:torque}).

Radio observations provided accurate measurements of the spindown torque
acting on the star (Camilo et al. 2007). The torque was decreasing together 
with the X-ray luminosity 2-3 years after the starquake. Its history at 
earlier times was not observed; the torque is believed to have increased 
after the starquake (Camilo et al. 2007). Such a non-monotonic evolution 
of the torque would be consistent with the theoretical expectations 
(\S~\ref{sc:torque}).


\section{Discussion}

Electrodynamics of untwisting may be summarized as follows:

1. --- The twist evolution following a starquake is not diffusive spreading 
that was pictured previously. Instead, the twist current is sucked into the 
star, a cavity immediately forms in the inner magnetosphere and grows 
until it erases all of the twist (Fig.~6). A sharp, step-like drop in current 
density $j$ is maintained at the boundary of the cavity. It is caused by the 
threshold nature of the discharge that conducts magnetospheric currents.

2. --- As the cavity expands and the j-bundle shrinks toward 
the magnetic axis, the twist amplitude $\ta$ in the j-bundle can grow.
The growth occurs if $d\V/du>0$, i.e. the discharge voltage is smaller on 
field lines extending farther from the star.
This is plausible if the current is conducted through $e^\pm$ 
discharge.\footnote{In contrast, if the current-carrying charges were 
   lifted from the star by voltage (\ref{eq:Vei}), the voltage would be 
   larger for field lines that extend to larger altitudes, i.e. $dV/du<0$. 
   Then the twist on the j-bundle would diminish with time rather than grow.}
The j-bundle with the growing twist is shrinking faster with time, so that 
the total twist energy $\Etor$ decreases consistently with the Ohmic 
dissipation rate. 

3. --- The growing twist that has reached the threshold for instability
$\tamax={\cal O}(1)$ is expected to ``boil over'' and drive an intermittent 
outflow of magnetic energy from the star. The twist in the j-bundle then
remains near $\tamax$ for the rest of its lifetime. It may be regulated by 
the limit-cycle instability --- the repeated growth of $\ta$ to $\tamax$ 
followed by a sudden reduction of $\ta$ below $\tamax$. The value of 
$\tamax$ and the nonlinear behavior of the outer magnetosphere at 
$\ta\sim\tamax$ needs to be studied further using numerical simulations.

4. --- In addition to rare large starquakes, the magnetosphere can be 
gradually twisted by the continued motion of its footpoints, either plastic 
or through a sequence of small starquakes. The continued footpoint motion 
leads to either a very strong twist 
or a tiny (negligible) twist that has no observational effects. 
A quasi-steady state with a non-negligible twist amplitude is possible 
only with $\ta\sim\tamax$. 

Recent observations of \XTE~ are particularly useful for testing the 
untwisting theory, for a few reasons:
(1) \XTE~ displayed a clean post-starquake evolution, which was observed 
for years uninterrupted by new starquakes. (2) The low level of quiescent 
luminosity from this object allows one to see clearly the shrinking hot spot 
on the star, and its evolving luminosity was tracked from $10^{35}$~erg/s 
down to $10^{33}$~erg/s. 
(3) The detection of radio pulsations and detailed measurements of spindown
torque make this object yet more interesting for testing theoretical models. 
For these reasons, this paper focused mainly on \XTE (\S~7).
Recently, an outburst was detected in the similar radio magnetar \AXP~ 
(Camilo et al. 2008; Halpern et al. 2008). Its behavior appears to be
more complicated than that of \XTE; apparently, episodes of repeated 
(and overlapping in time) activity occurred a few months after the outburst. 
The analysis of this object is deferred to a future work.

Other, more active, AXPs and SGRs display a diverse and complicated behavior 
of X-ray luminosity, pulse profile, and spindown rate (see Woods \& Thompson
2006; Kaspi 2007; Mereghetti 2008 for reviews). Repeated starquakes
of various amplitudes and possible plastic deformations of the crust 
make these objects more difficult to analyze. The continuing injection of 
a magnetospheric twist can slow down the decay of its luminosity. 
It can also affect the star's spindown in a more 
complicated way than described in \S~\ref{sc:torque}
for an isolated starquake. In general, twist injection should
lead to higher X-ray activity and faster spindown (Thompson et al. 2002). 
Such a general correlation exists in the magnetar population
(Marsden \& White 2001) but not always observed in individual objects.
Note that spindown is controlled by the X-ray-dim bundle of open field 
lines. The impact of a starquake on this narrow bundle may be immediate, 
occur with a delay, or never occur, depending on the starquake geometry 
and amplitude (cf. the case of a ``ring'' starquake in \S~\ref{sc:ring}). 
Repeated starquakes may lead to a non-trivial relation between 
spindown and X-ray emission. 

Despite the diverse behavior of active magnetars, some general features
may be inferred. It is clear that the observed sources do not have strong 
global twists, for two reasons. First, the evolution timescale of such 
twists would be too long, $\tev\sim(10-10^2)\V_9^{-1}$~yr, where $\V_9$ 
is the discharge voltage in units of $10^9$~V (see eq.~\ref{eq:tev}). 
In contrast, the magnetospheres of observed magnetars usually evolve on 
timescales $\sim 1$~yr or even shorter.\footnote{The theoretical $\tev$ would 
    be reduced for a larger discharge voltage $\V\gg 10^9$~V, however it 
    appears impossible to sustain such a voltage as it leads to runaway 
    $e^\pm$ creation (BT07). Even if $\V\gg 10^9$~V were theoretically 
    possible, strong global twists would still be ruled out by observations 
    because they are overluminous.}
Second, the luminosity produced by a strong global twist would be too high,
$L\sim 10^{37}\V_9$~erg~s$^{-1}$ (see eq.~\ref{eq:lum}). It is 2 orders 
of magnitude higher than the typical observed luminosities of magnetars,
$L\sim 10^{35}$~erg/s (e.g. Durant \& van Kerkwijk 2006). This leaves two 
possibilities: the twist is weak ($\ta\ll 1$) or localized to a narrow 
bundle of field lines. The changing spindown rate suggests a strong twist, 
at least near the magnetic dipole axis. Thus, observations appear to 
support the picture of a strongly twisted j-bundle near the dipole axis.
It is certainly supported by the observations of \XTE~ (\S~\ref{sc:XTE}). 

The localized strong twist may be created by starquakes localized to 
the polar region. It also tends to form dynamically from a weak global 
twist as the j-bundle shrinks to the axis.
The luminosity and evolution timescale of an untwisting magnetosphere  
with $\uf=\sin^2\theta_\star\sim 0.1$ and $\ta\sim 1$ is consistent with  
typical $L\sim 10^{35}$~erg/s and 
$\tev\sim 1$~yr of active magnetars (see eqs.~\ref{eq:lum} and \ref{eq:tev}).
Both nonthermal X-ray components in the magnetar spectra, 1-20~keV and 
20-300~keV, can be produced by the j-bundle.
They are likely emitted at different radii. 
If the hard component is produced near the star, where the j-bundle 
is narrow, a relatively narrow 20-300~keV pulse is expected. 

The resonant-scattering model for 1-20~keV radiation is consistent with the 
picture of a narrow j-bundle near the dipole axis. The cyclotron resonance 
for $e^\pm$ with keV photons takes place at radii $r\sim 10R$. This means 
that the resonant scattering is confined to the field-line bundle with 
$u=R/\Rmax\sim 0.1$ (\S~\ref{sc:nonth}). The luminosity of upscattered 
radiation $L_{\rm sc}$ is supplied by Ohmic dissipation of electric
currents in this bundle. 

The footprints of a bundle with $u\sim 0.1$ form $\sim 3$~km spot on the star. 
A significant fraction of energy released in the j-bundle may be transported 
to its footprints and emitted there quasi-thermally.
If the spot radiates a significant part of the bundle luminosity, 
$L\sim 10^{35}$~erg/s, then it must have a temperature $kT\approx 1$~keV. 
Such hot spots were observed in \XTE~ and \AXP,
but were not reported for other, brighter magnetars.
The typical reported temperatures of blackbody components are
$0.3-0.6$~keV (e.g. Perna et al. 2001).
The disappearance of hot spots in the presence of large $L_{\rm sc}$
may be caused by the outward drag created by resonant scattering,
which suppresses the transport of released energy to the footprints of the 
j-bundle.

\acknowledgments
This work was supported by NASA grant NNG-06-G107G.


\section*{Appendix A: Twisted dipole magnetosphere} 

Consider a weakly twisted dipole field $\bB=\bB_0+\bB_\phi$.
Its poloidal component $\bB_0$ is that of untwisted dipole,
\be
\label{eq:B0}
  B_r=\frac{2\mu\cos\theta}{r^3}, \qquad B_\theta=\frac{\mu\sin\theta}{r^3}.
\ee
The poloidal flux function is given by
\be
\label{eq:fm_app}
   \fm(r,\theta)=\int_0^\theta\int_0^{2\pi} B_r h_\phi h_\theta\, d\theta\,d\phi
                =2\pi\mu\,\frac{\sin^2\theta}{r},
\ee
where $h_\phi=r\sin\theta$ and $h_\theta=r$. 
Since $f=const$ on any flux surface, 
\be
\label{eq:Rmax_app}
  \frac{r}{\sin^2\theta}=const=\Rmax,
\ee 
which is the radius where the flux surface crosses the equatorial plane
(the maximum radius reached by the flux surface). 

The magnitude of the twisted magnetic field is given by 
(neglecting terms ${\cal O}[B_\phi^2/B^2$]),
\be
   B(r,\theta)\approx B_0=\frac{\mu}{r^3}\,\sqrt{1+3\cos^2\theta}.
\ee
The twist angle $\ta$ (eq.~\ref{eq:ta}) is easy to calculate using 
$\theta$ as the parameter along the field line and substituting 
$dl=(B/B_\theta)d\theta$, 
\be
  \ta=\int_{\theta_0}^{\pi-\theta_0} \frac{B_\phi}{B_\theta}
      \frac{h_\theta d\theta}{h_\phi}.
\ee
Here $\theta_0$ is the polar angle of the northern footpoint of the field
line, and $\pi-\theta_0$ is the polar angle of the southern footpoint. 
Substituting $B_\phi=2I/ch_\phi$ (eq.~\ref{eq:Bphi}), $B_\theta$ from
equation~(\ref{eq:B0}), and using equation~(\ref{eq:Rmax_app}), one finds
\be
\label{eq:ta1}
  \ta=\frac{4I}{c\mu}\,\Rmax^2\cos\theta_0,
\ee
where 
\be
     \cos^2\theta_0=1-\frac{\rc}{\Rmax}=1-\frac{\fm}{\fmc}.
\ee
When the poloidal current $I$ and twist angle $\ta$ are viewed as  
functions of $u=\fm/\fms$ and time $t$, the relation between $\ta$ and 
$I$ (eq.~\ref{eq:ta1}) becomes equation~(\ref{eq:ta2}).
The free energy of a twist with given $I(u)$ is found using
equations~(\ref{eq:en0}) and (\ref{eq:ta2}),
\be
\label{eq:Etw_app}
   \Etor(t)=\frac{2R}{c^2}\int_0^{u_c}I^2(u,t)\sqrt{1-\frac{u}{u_c}}
           \,\frac{du}{u^2}.
\ee

We also calculate here the integral that is needed in \S~5.1,
\be
 \int_{r>R}B\,dl\approx \int_{r>R}B_0\,dl.
\ee
It is taken along the field line outside the star.
Consider the contour that closes the path of integration between the 
two footpoints along the surface of the star (and across the magnetic field). 
The line integral of $\bB_0$ along this closed contour vanishes (as follows 
from the Stokes' theorem and $\nabla\times \bB_0=0$). This gives
\be
\label{eq:int}
  \int_{r>R} B_0\,dl= \int^{\pi-\theta_1}_{\theta_1} B_\theta Rd\theta
                     =\frac{2\mu\cos\theta_1}{R^2}.
\ee
Here $\theta_1$ and $\pi-\theta_1$ are the polar angles of the field-line
footprints on the star surface, and $\cos\theta_1=\sqrt{1-u}$.

\section*{Appendix B: Derivation of the front equation}

The front is located on the flux surface $\uf$ where $\Phi_e(u)$ deviates 
from $\V(u)$, current $j$ jumps to $\sim\js$ (Fig.~5). We consider below 
the limit $\js\rightarrow 0$. Then the jump of $j$ is described by the 
Heaviside step function $\Theta(\uf-u)$ and $j=0$ in the region $\uf<u<1$.

This implies that $I(u)=const$ for $\uf<u<1$, and this fact can be used
to derive the dynamical equation for $\uf$. Let us differentiate the 
twist evolution equation~(\ref{eq:evd2}) with respect to $u$. Then 
the left-hand side vanishes in the region $\uf<u<1$ and we get
\be
 \frac{\partial}{\partial u}\left(\frac{c^2u^2}{4R\sqrt{1-u/\uc}}\,
   \frac{\partial\Phi_e}{\partial u}\right)=0, \qquad \uf<u<1,
\ee
which implies
\be
   \frac{\partial\Phi_e}{\partial u}=\frac{K}{u^2}\sqrt{1-\frac{u}{\uc}},
\ee
where $K$ is constant in the region $\uf<u<1$. Integrating this equation and 
using the boundary condition $\Phi_e(1)=0$ we find the shape of $\Phi_e(u)$,
\be
\label{eq:Phi}
   \Phi_e(u)=\left\{ \begin{array}{ll}
     \V(u)  &  0<u<\uf \\ 
 \displaystyle -\frac{K}{\uc}\left[-\frac{\sqrt{1-x}}{x}
 +\frac{1}{2}\ln\frac{1+\sqrt{1-x}}{1-\sqrt{1-x}}\right]_{x=u/\uc}^{x=1/\uc}
        &  \uf<u<1 \\ 
                     \end{array}
            \right.
\ee 
The explicit expression for $K$ can be found from the condition 
$\Phi_e(\uf)=\V(\uf)$, 
\be
\label{eq:K}
  K(\uf) = -\V(\uf)\,\uf\,\xi(\uf), 
\ee
\be
\label{eq:xi}
  \xi   =  \frac{\uc}{\uf}\,\left[\frac{\uc}{\uf}\,\sqrt{1-\frac{\uf}{\uc}}
   -\uc\sqrt{1-\frac{1}{\uc}}+\frac{1}{2}\,\ln\left(
   \frac{1+\sqrt{1-\uc^{-1}}}{1-\sqrt{1-\uc^{-1}}}\,\cdot\,
   \frac{1-\sqrt{1-\uf/\uc}} {1+\sqrt{1-\uf/\uc}} \right)\right]^{-1}.
\ee
The numerical factor $\xi(\uf)$ is well approximated by a simpler formula,
\be
\label{eq:xi1}
   \xi\approx \frac{1+1.4\uf}{1-\uf}.
\ee
The accuracy of this approximation is better than 2.2\% for $0<\uf<0.65$.
The approximation is worst (30\% accuracy) when $\uf=1$ and quickly becomes
excellent as the front propagates away from the boundary.

The twist evolution equation~(\ref{eq:evd2}) can now be written as
\be
\label{eq:dIdt}
  \frac{\partial I}{\partial t}=\left\{ \begin{array}{ll}
   \displaystyle \frac{c^2u^2 \V^\prime(u)}{4R\sqrt{1-u/\uc}} & 0<u<\uf \\ 
   \displaystyle \frac{c^2K}{4R}                              & \uf<u<1 \\ 
                                         \end{array}
                                \right. 
\ee 
where $\V^\prime=d\V/du$. 
Note that $\partial I/\partial t$ does not vary with time at $u<\uf$, 
and hence $I(u,t)$ can be obtained by simple integration. 
Thus we find the current function of the twist,
\be
\label{eq:I_app}
   I(u,t)=\left\{ \begin{array}{ll} \displaystyle 
   I_0(u)+\frac{c^2u^2 \V^\prime(u)}{4R\sqrt{1-u/\uc}}\,t &  0<u<\uf \\
   \If(t) &  \uf<u<1 \\
                  \end{array}
          \right. 
\ee
where $I_0(u)\equiv I(u,0)$ is the initial current function, and 
$\If(t)\equiv I[\uf(t),t]$ is the net current that flows through 
the magnetosphere at time $t$. Using the relation between $I$ and $\ta$
(eq.~\ref{eq:ta2}), we obtain equation~(\ref{eq:psi}). It gives a 
complete solution to the problem of twist evolution if we find $\uf(t)$.

The equation for $\uf(t)$ can be obtained if one considers the evolution 
of $\If$. Its time derivative is given by
\be
  \frac{d\If}{dt}=\frac{\partial I}{\partial t}\left(\uf,t\right)
        +\frac{\partial I}{\partial u}\left(\uf,t\right)\,\frac{d\uf}{dt},
\ee
which can be evaluated separately on each side of the front, 
\be
  \left.\frac{d\If}{dt}\right|_{\uf-0}
  =\frac{c^2\uf^2 \V^\prime(\uf)}{4R\sqrt{1-\uf/\uc}}
  +\left(\left.\frac{dI_0}{du}\right|_{\uf}
  +\left.\frac{d}{du}\frac{c^2u^2\V^\prime(u)}
   {4R\sqrt{1-u/\uc}}\right|_{\uf}\,t\right) \frac{d\uf}{dt},
\ee
\be
  \left.\frac{d\If}{dt}\right|_{\uf+0}=\frac{c^2K}{4R}. 
\ee
The two results must match because they describe the same $\If(t)$.
The jump in $\partial I/\partial u$ across the front (which corresponds to 
the jump in $j$) is compensated by the jump in $\partial I/\partial t$. 
Equating the two expressions for $d\If/dt$, we find
\be
\label{eq:front_app}
\displaystyle
   \frac{d\uf}{dt}=\frac{\displaystyle K(\uf)-\frac{\uf^2\V^\prime(\uf)}
                                         {\sqrt{1-\uf/\uc}}}
     {\displaystyle \frac{4R}{c^2}\left.\frac{dI_0}{du}\right|_{\uf}
  +\left.\frac{d}{du}\frac{u^2\V^\prime(u)}{\sqrt{1-u/\uc}}\right|_{\uf} t}.
\ee 
This ordinary differential equation describes the propagation of the front. 
Substitution of equation~(\ref{eq:K}) for $K(\uf)$ gives 
equation~(\ref{eq:front}).


\end{document}